\documentclass[pra,twocolumn,superscriptaddress,notitlepage]{revtex4}

\usepackage{graphicx}
\usepackage{color}
\usepackage{amssymb}
\usepackage{amsmath}
\usepackage{bm}

\usepackage[normalem]{ulem} 

\definecolor{darkblue}{rgb}{0.,0.24,0.51}
\definecolor{britishracinggreen}{rgb}{0.0, 0.26, 0.15}
\definecolor{darkgreen}{rgb}{0,0.60,.2}

\usepackage[breaklinks=true,colorlinks=true,linkcolor=darkblue,urlcolor=darkblue,citecolor=darkblue]{hyperref}

\newcommand{\down}{\downarrow}
\newcommand{\up}{\uparrow}
\renewcommand{\k}{{\bf k}}
\newcommand{\p}{{\bf p}}
\newcommand{\q}{{\bf q}}
\newcommand{\0}{{\bf 0}}

\renewcommand{\r}{{\bf r}}
\newcommand{\ef}{\varepsilon_F}

\newcommand{\kf}{k_F}

\newcommand{\aho}{a_{\rm{ho}}}

\begin{document}

\title{Universality of the unitary Fermi gas: A few-body perspective}

\author{Jesper Levinsen}
\address{School of Physics and Astronomy, Monash University, Victoria 3800, Australia}

\author{Pietro Massignan}
\address{ICFO-Institut de Ciencies Fotoniques, The Barcelona Institute of Science and Technology, 08860 Castelldefels, Spain}

\author{Shimpei Endo}
\address{School of Physics and Astronomy, Monash University, Victoria 3800, Australia}

\author{Meera M.~Parish}
\address{School of Physics and Astronomy, Monash University, Victoria 3800, Australia}
\email{meera.parish@monash.edu}

\begin{abstract}
  We revisit the properties of the two-component Fermi gas with
  short-range interactions in three dimensions, in the limit where the
  $s$-wave scattering length diverges. Such a unitary Fermi gas
  possesses universal thermodynamic and dynamical observables that are
  independent of any interaction length scale.  Focusing on trapped
  systems of $N$ fermions, where $N\leq10$, we investigate how well we
  can determine the zero-temperature behavior of the many-body system
  from published few-body data on the ground-state energy and the
  contact.  For the unpolarized case, we find that the Bertsch
  parameters extracted from trapped few-body systems all lie within
  15\% of the established value.  Furthermore, the few-body values for
  the contact are well within the range of values determined in the
  literature for the many-body system.  In the limit of large spin
  polarization, we obtain a similar accuracy for the polaron energy,
  and we estimate the polaron's effective mass from the dependence of
  its energy on $N$.  We also compute an upper bound for the squared
  wave-function overlap between the unitary Fermi system and the
  non-interacting ground state, both for the trapped and uniform
  cases. This allows us to prove that the trapped unpolarized ground
  state at unitarity has zero overlap with its non-interacting
  counterpart in the many-body limit $N \to \infty$.
\end{abstract}

\maketitle

\vspace{2pc}

\section{Introduction}

Dilute gases of fermions with short-range interactions and diverging
scattering lengths are of relevance to neutron stars, ultracold atomic
gases, and possibly even to unconventional superconductors in the
solid state \cite{Giorgini2008RMP,parish2015}.  Of particular interest
is the two-component ($\up$, $\down$) Fermi system, which has been
shown to be stable in the unitarity limit where the $s$-wave
scattering length $a \to \pm \infty$
\cite{Heiselberg2012,PhysRevA.67.010703,PhysRevLett.93.090404}.  In
this case, there is no interaction length scale at low energies, and
the unitary Fermi gas displays a universal equation of state that only
depends on the particle density, the spin polarization, and the
temperature~\cite{Ho2004,Navon2010,Horikoshi2010,Ku2012,van2012feynman}.
Moreover, the system is spatially scale invariant, which results in
universal properties such as a vanishing bulk viscosity
\cite{son2007vis,Castin2012,Cao2011}.

Despite its symmetry properties, the unitary many-body system is
notoriously difficult to treat theoretically, since there is no small
interaction parameter.  Thus, a variety of sophisticated numerical
methods have been used to attack the problem.  The ground-state and
finite-temperature thermodynamics of the unpolarized gas have been
investigated using pairing fluctuations approaches, density-functional
theory, and various types of quantum Monte Carlo (QMC)
\cite{Hammond1994book,Randeria2012,Strinati2012,Bulgac2012}.  In the
limit of large spin imbalance, where the system is well described by
Fermi liquid theory, successful theoretical approaches include
fixed-node and diagrammatic QMC, variational wave functions, and
ladder diagrams \cite{Chevy2010,Massignan2014review}.

Given that the unitary Fermi gas lacks a small interaction parameter,
it is of interest to consider other perturbative approaches.  A
prominent example is the virial expansion, which is applicable in the
regime where the system's temperature $T$ is far above quantum
degeneracy.  Taking advantage of the fact that the fugacity
$z=e^{\mu/k_BT}$ is small in this limit, the entire thermodynamics may
be determined from an expansion of the thermodynamic potential
\begin{align*}
\Omega=-k_BT\frac{2V}{\lambda^3}\sum_{N\geq1}b_Nz^N.
\end{align*}
Here, $V$ is the volume, $k_B$ the Boltzmann constant, $\lambda$ the
thermal wavelength, and $\mu$ the chemical potential, where we have
considered the unpolarized system for concreteness.  Crucially, the
thermodynamic potential depends only on {\em few-body correlations}
through the virial coefficients $b_N$
\cite{Liu2009,Rakshit2012,Yan2016,Endo2016}, which are completely
determined from the energy spectrum of $N$ particles.  Recent
cold-atom experiments on the thermodynamics of the unitary Fermi gas
\cite{Nascimbene2010,Ku2012} have observed very good agreement with
the results from the virial expansion, even for a weakly degenerate
gas.  For a review of the virial expansion, we refer the reader to
Ref.~\cite{Liu2013}.

In this paper, we consider the opposite limit and ask whether few-body
physics can also be used to gain insight into the behavior of a
strongly interacting unitary Fermi gas at \textit{zero temperature}.
Our approach is particularly motivated by recent work on the
harmonically trapped one-dimensional Fermi gas. Here, the Heidelberg
experiment led by S.~Jochim explored the evolution from few- to
many-body physics using a single $\down$ impurity atom repulsively
interacting with $N_\uparrow$ spin-$\up$ fermions, where $N_\uparrow$
was increased from 1 to 5~\cite{Wenz2013}. In this few-body system,
they observed a fast convergence of the interaction energy towards the
expected many-body result~\cite{Astrakharchik2013}. Subsequently, it
has been predicted that both the energy and contact of a spin-balanced
gas in a one-dimensional harmonic trap converge rapidly towards the
many-body limit, for any interaction strength \cite{Grining2015}.
Thus, this raises the question of whether few-body correlations can
similarly determine the behavior of the three-dimensional unitary
Fermi gas.

To investigate the relationship between the few- and many-body systems
at unitarity, we analyse data from trapped systems of $N$ equal-mass
fermions, where $N \leq 10$. Unless otherwise specified, the results
we quote arise from calculations for the unitary harmonically trapped
Fermi gas. For a comprehensive review of few-body physics in harmonic
traps, we refer the reader to Ref.~\cite{Blume2012}.  We focus on
static quantities at zero temperature such as the energy, the contact,
and the squared wave-function overlap with the non-interacting state.
Furthermore, we only consider the ground state of the system and we
ignore the behavior of the metastable repulsive branch, such as the
possibility of itinerant ferromagnetism \cite{Valtolina2016}.

We emphasize that our primary objective is not to provide precise
predictions for the many-body properties of the unitary Fermi
gas. Rather, we will determine which of these properties may be
reasonably well described within few-body calculations, and whether
these properties converge quickly towards the many-body limit.  We
also discuss how the squared wave-function overlap between the unitary
and non-interacting ground states evolves with particle number $N$,
and we calculate an upper bound for this overlap for general $N$.

The manuscript is organized as follows.  In Section \ref{sec:UFG} we
review the fundamental properties of the uniform unitary Fermi gas,
focussing on equal spin populations and large spin imbalance, and we
introduce the local density approximation for trapped systems of
particle number $N \gg 1$.  In Section \ref{sec:energy} we show how
one may obtain surprisingly accurate predictions for the energy of the
unitary Fermi gas in the thermodynamic limit from a knowledge of the
energies of very few trapped atoms ($2\leq N \leq 10$).  The related
analysis presented in Section \ref{sec:contact} once again leads to an
extrapolated contact which lies well within the range of the most
recent theoretical and experimental determinations.  In Section
\ref{sec:residue} we consider the overlaps between the non-interacting
and strongly interacting wave functions.  Here we derive a rigorous
upper bound of this overlap and we determine a general relationship
between the overlap in the trapped system and that in uniform space.
We conclude in Section \ref{sec:conclusions}, by presenting an
outlook, and by summarizing our main results.

\section{The unitary Fermi gas}
\label{sec:UFG}

We begin our study of the unitary Fermi gas by considering the
universal properties of the many-body system with spin components
$\sigma = \up, \down$, where the total number of fermions
$N = N_\up + N_\down \gg 1$.  To simplify the discussion, we focus on
the equal-mass case, where $m_\up = m_\down \equiv m$, and we work in
units where $\hbar = k_B = 1$.

For a uniform system at temperature $T$ and in a volume $V$, the
general equation of state for the total energy $E$ can be written as:
\begin{equation}
\label{eq:EoS}
E= \frac{3}{5} \frac{k_F^2}{2m} N \;
\beta\left(\frac{T}{E_F},
  {\frac{N_\down}{N_\up}},
  \frac{1}{k_F a}
\right) .
\end{equation}
Here, we assume that $N_\down \leq N_\up$, without loss of generality,
and we define the Fermi momentum $k_F = (6\pi^2 N_\up/V)^{1/3}$ and
Fermi energy $E_F = \kf^2/2m$.  We neglect the microscopic details of
the interaction, such as the range $r_0$, since we are considering the
regime of low-energy resonant scattering, $k_F |r_0| \ll 1$ and
$|r_0| \ll |a|$.

Taking the limits $1/a \to 0$ and $T \to 0$, it is clear that
Eq.~\eqref{eq:EoS} becomes a universal function of the particle
numbers $N_\up$, $N_\down$.  In particular, when the system is
unpolarized, i.e., $N_\up = N_\down$, we obtain the universal Bertsch
parameter: $\xi \equiv \beta(0,1,0)$~\cite{Heiselberg2012}. In the
opposite limit of a large spin polarization, $N_\down/N_\up \to 0$, we
have $\beta \to 1$.

One can fully characterize the universal thermodynamics of the uniform
unitary Fermi gas at zero temperature from just three observables: the
energy $E$, the contact
\begin{align}
            C = - 4\pi m \left(\frac{\partial E}{\partial a^{-1}} \right)_{N_\up,N_\down} ,
            \label{eq:C}
\end{align}
and the (inverse) spin susceptibility 
\begin{align}\label{inverseSpinSusceptibility}
\chi_s^{-1} =  V \left(\frac{\partial^2 E}{\partial M^2}\right)_N ,
\end{align}
where the magnetization $M = N_\up - N_\down$. In principle, there are
other static observables such as the compressibility $\kappa$ and the
heat capacity $\mathcal{K}_V$, but $\mathcal{K}_V \to 0$ in the limit
of zero temperature, while the compressibility is simply related to
the energy: $\kappa^{-1} = \frac{10}{9} \frac{E}{V}$.  We will not
concern ourselves with dynamical (transport) quantities like the
viscosity~\cite{son2007vis,Castin2012,Cao2011} or conductivity. These
also have interesting symmetry properties, but are non-trivial to
extract from the few-body system.

\subsection{Unpolarized gas}
\label{sec:unpol} 

The behaviour of the unitary Fermi gas at zero temperature simplifies
considerably when $N_\up = N_\down$. In this case, the energy in
Eq.~\eqref{eq:EoS} reduces to
\begin{align} \label{eq:unpol}
    E = \xi \frac{3}{5} \frac{k_F^2}{2m} N \equiv \xi E_0 ,
\end{align}
where $E_0$ is the energy of the non-interacting system.  Thus, the
thermodynamic quantities that directly follow from
Eq.~\eqref{eq:unpol} only depend on the particle density and the
Bertsch parameter.  Since the ground state of the unpolarized system
is a paired superfluid, another quantity of interest is the pairing
gap $\Delta$. This is the minimum energy required to add an unbound
quasiparticle and is thus a feature of the quasiparticle excitation
spectrum. The existence of such a gap results in an exponentially
vanishing spin susceptibility as $T \to 0$ \cite{Chevy2012}.

However, it is generally not straightforward to relate the pairing gap
in the trapped gas to that in the uniform system. While one can
define an excitation gap for having an unbound particle in the trapped
system \cite{Blume2012}, this extra particle becomes confined to the
trap edge as the number of particles $N \to \infty$. Thus, it will
sensitively depend on the system boundary and it cannot be considered
a ``bulk'' property. Indeed, even experiments on trapped ultracold
atomic gases must perform local radio-frequency spectroscopy near the
trap center in order to extract the pairing gap for a uniform system
\cite{Schirotzek2008}.

To define a bulk pairing gap for the trapped gas, one can analyse the
spatially varying pairing field within mean-field theory
\cite{Bruun2002}. This analysis would suggest that we have the local
contact density $C/V = m^2 (\Delta^2 + U^2)$ at every point in the
trap, where $U$ is the Hartree energy shift in the quasiparticle
spectrum \cite{Leggett1980}. Using the experimental observation that
$U \approx \Delta$ at unitarity \cite{Schirotzek2008}, we then obtain
$C/V \approx 2m^2 \Delta^2$.  Therefore, we will use the contact as a
measure of the pairing gap in the trapped unitary system.

\subsection{Limit of large spin polarization}
\label{sec:chevy}

When there is a small concentration of spin-$\down$ atoms,
$x=N_\down/N_\up\ll1$, the system can be described in terms of dressed
impurities, or ``polarons", immersed in an ideal Fermi sea.  Such
polarons are weakly interacting quasiparticles characterised by three
parameters: the impurity energy $E_{\rm{pol}}$ at zero momentum, the
effective mass $m^*$, and the residue $Z$ (defined as the squared
overlap between the interacting and non-interacting ground-state wave
functions). For a comprehensive survey of the vast literature on
Fermi polarons, we refer the reader to recent reviews covering the
subject \cite{Chevy2010,Massignan2014review,Levinsen2015review}.

These basic quasiparticle properties define the zero-temperature
equation of state for the polarized Fermi gas, which takes the
Landau-Pomeranchuk form \cite{Lobo2006, Pilati2008}
\begin{align}
\label{energyFunctional}
E=\frac{3}{5}E_F N_\up+\left(E_{\rm pol} + \frac{3}{5}E_F \frac{m}{m^*}x^{2/3}
\right) N_\down
\end{align}
in the limit $x\ll1$.  The terms on the right hand side of
Eq.~\eqref{energyFunctional} correspond to, respectively, the energy
of the unperturbed majority Fermi sea, the energy of the impurities
dressed by the majority atoms, and the Fermi pressure associated with
the finite concentration of dressed impurities. At this level of
approximation, the polarons behave as ideal quasiparticles, since the
interactions between polarons contribute only at order $x$ to the
system's energy~\cite{Pilati2008}.

From Eqs.~\eqref{inverseSpinSusceptibility} and
\eqref{energyFunctional}, the spin susceptibility in this limit is
\begin{align}
\chi =  x^{1/3} \frac{6 m^*}{E_F m} \frac{N_\up}{V}
\equiv \frac{m^*}{m}\chi_0,
\end{align}
where $\chi_0$ is the spin susceptibility of the non-interacting Fermi
gas in the ground state.  Thus, $\chi$ provides a direct probe of the
effective mass of the polaron.
 
In a uniform system, the quasiparticle properties can be accurately
described within a simple variational approach \cite{Chevy2006}, where
only a single particle-hole excitation of the Fermi sea is included.
Specifically, for a polaron at zero momentum, the wave function within
the one-particle-hole (1PH) Ansatz reads
\begin{align}\label{1PH_ansatz}
|\psi_\0\rangle= \phi_0 c^{\dag}_{\0\down}|FS\rangle_\up 
+ \sum_{q<\kf}^{k>\kf}\phi_{\k \q}c^{\dag} _{\q-\k_\down}c^{\dag}_{\k\up}\,c_{\q\up}\,|FS\rangle_\up,
\end{align}
where $c_{\p\sigma}$ annihilates a particle with spin $\sigma$ and
momentum $\p$, $|FS\rangle_\up$ is the ground-state of the
non-interacting spin-$\up$ Fermi sea, and $\phi_0, \phi_{\k\q}$ are
variational parameters. Minimizing the total energy of the system
leads to a self-consistent equation of the form
$E_{\rm pol}=\Sigma(\p=\0,E_{\rm pol})$ \cite{Chevy2006}, which is
formally equivalent to finding the real pole of the dressed impurity
Green's function \cite{Combescot2007}
\begin{align}
G_\down(\p,\omega)=\frac1{\omega -\varepsilon_{\p}-\Sigma(\p,\omega)+i0_+}.
\end{align}
Here $\varepsilon_{\p}=p^2/2m$, and $\Sigma$ is the retarded impurity
self-energy computed within the T-matrix (or ``forward-scattering")
approximation, which at zero momentum is given by
\begin{align}\label{self_energy}
  &\Sigma(\0,\omega)
=\sum_{q<\kf}\left[
\frac{m}{4\pi a}\right. \nonumber\\
&\left.-\sum_{k>\kf}\left(\frac{1}{\omega-(\varepsilon_{\mathbf{q-k}}+\varepsilon_{k}-\varepsilon_{q})+i0_+}+\frac{m}{k^{2}}\right)\right]^{-1}.
\end{align}
Using this approach, one finds $E_{\rm{pol}}\approx-0.61E_F$,
$Z\approx0.78$, and $m^*\approx1.17m$. These values compare remarkably
well with the results of state-of-the-art calculations:
$E_{\rm{pol}}=-0.615E_F$, $Z=0.759$, and
$m^*=1.197m$~\cite{Prokofev2008,Combescot2008,Vlietinck2013}.

The fact that one obtains such excellent results from a simple Ansatz,
which only accounts for \textit{two-body} correlations exactly,
motivates us to analyze the quasiparticle properties from the few-body
limit.

\subsection{Trapped system and the local density approximation}

To connect with the trapped few-body system, we must consider the
Fermi gas in the presence of an isotropic harmonic potential
$V(\r) = \frac{1}{2}m\omega^2 r^2$, where $\omega$ is the trapping
frequency.  We implicitly assume that the harmonic oscillator length
$\aho \equiv 1/\sqrt{m\omega}$ is much larger than the range $|r_0|$,
so that we remain in the unitary regime.

In the limit where the total number of particles $N \gg 1$, the local
density approximation (LDA) holds \cite{PethickSmith2008book} and we
may determine global quantities from averages over local quantities in
the trap, i.e., for a local observable $F$, we have trapped observable
$\mathcal{F} = \frac{1}{V}\int d^3r F(\mu(\r))$ \footnote{In general,
  we will use calligraphic symbols to denote the trapped thermodynamic
  observables.}.  Here, we define the local chemical potential
$\mu(\r) = \mu - V(\r)$, which is connected to the global chemical
potential
 \begin{align} \label{eq:globchem}
     \mu = \frac{\partial \mathcal{E}}{\partial N},
 \end{align}
where $\mathcal{E}$ is the total energy of the trapped gas. 

For a trapped, non-interacting Fermi gas with balanced spin
populations, we have particle density
\begin{align} \label{eq:dens}
    n(\r) = \frac{1}{3\pi^2} (2 m)^{3/2} \left( \mu - \frac{1}{2} m\omega^2 r^2  \right)^{3/2},
\end{align}
thus giving the total number of particles
\begin{align} \label{eq:num}
    N = N_\up + N_\down = 
    \int d^3r \,  n(\r) 
    = \frac{\mu^3}{3\omega^3} .
\end{align}
Similarly to the uniform system, we can use the relationship between
particle number and $\mu$ in Eq.~\eqref{eq:num} to define a Fermi
momentum and Fermi energy:
\begin{align}
    \kappa_F = (48N_\up)^{1/6}\aho^{-1}\, , & \hspace{8mm}
    \ef = \kappa_F^2/2m.
    \label{eq:ktoN}
\end{align}
These quantities are equivalent to the local Fermi momentum and Fermi
energy at the center of the trap.

From Eq.~\eqref{eq:globchem}, we find the energy of the
non-interacting trapped system to be
\begin{align} \label{eq:nonint}
\mathcal{E}_0 = \int^N_0 dN' \mu(N') 
= \frac{\omega}{4} (3N)^{4/3} = 2 \mathcal{E}_{0\up}
\end{align}
where the energy of the spin-$\up$ component is given by:
\begin{align}
\mathcal{E}_{0\up} = \frac{\omega}{8} (6N_\up)^{4/3} .
\end{align}
Note that one cannot directly average the energy over the trap, since
it is not a function of the local chemical potential. However, one can
perform a trap average on the thermodynamic potential
$\Omega(\mu,h) = E -\mu N - h M$, where $h$ is an effective Zeeman
field.

In the presence of interactions, the contact of the trapped system is
\cite{Blume2012}
\begin{align}
\mathcal{C}   = - 4\pi m \left(\frac{\partial \mathcal{E}}{\partial a^{-1}} \right)_{N_\up,N_\down},
\end{align}
which is equivalent to a trap averaged contact within LDA since we can
write
\begin{align}
  \mathcal{C} = - \frac{4\pi m}{V} \int d^3r \frac{\partial \Omega(\mu(\r),h)}{\partial a^{-1}} = 
  \int d^3r \, \frac{C(\mu(\r))}{V} \, .
\end{align}
Here, we have used the fact that the contact can be defined from
either $E$ or $\Omega$ \cite{Werner2012a}.

For the unitary unpolarized Fermi gas, we see from
Eq.~\eqref{eq:unpol} that we must simply replace $\mu(\r)$ by
$\xi^{-1}\mu(\r)$ in Eq.~\eqref{eq:dens}, thus giving
\begin{align}
N =    \frac{\mu^3}{3\omega^3\xi^{3/2}}\, , \ & \hspace{8mm} 
 \mathcal{E} = \sqrt{\xi} \, \mathcal{E}_{0}.
 \label{eq:xi1}
\end{align}
Furthermore, the dimensionless trap averaged contact
is~\cite{Werner2009,Hoinka2013}
\begin{align}
    \frac{\mathcal{C}}{N \kappa_F} = \frac{256}{105\pi \xi^{1/4}} \left(\frac{C}{Nk_F}\right),
    \label{eq:Crelation}
\end{align}
where the dimensionless contact for the unpolarized uniform system at
unitarity is given by
\begin{align}
    \frac{C}{Nk_F} = -\frac{6\pi}{5} \left.
\frac{\partial \beta(0,1,y)}{\partial y}\right|_{y=0}\, .
\end{align}

\section{Energy and the Bertsch parameter}
\label{sec:energy}
\begin{figure}[t]
\centering
\includegraphics[width=\columnwidth]{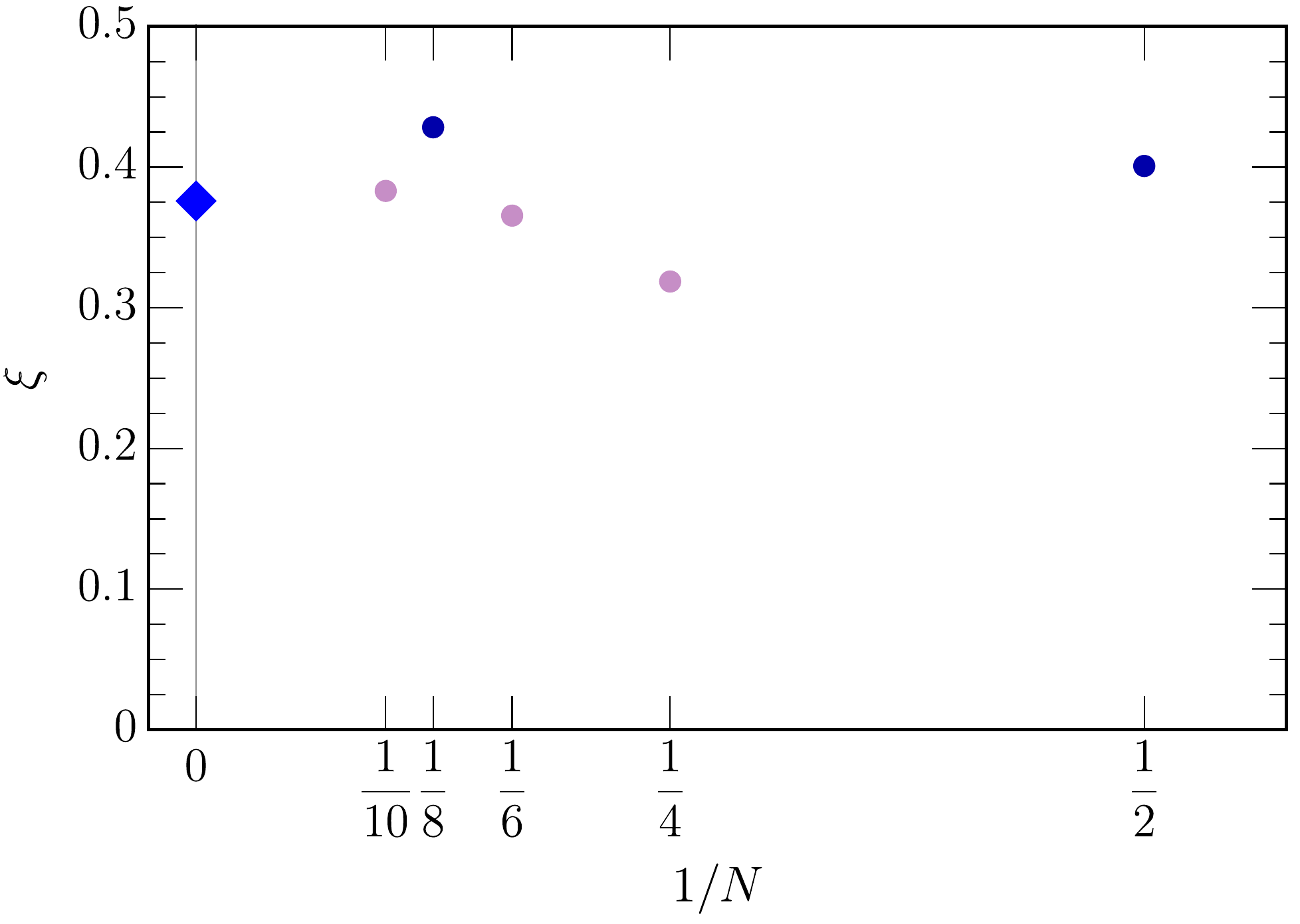}
\caption{\label{fig:bertsch} Bertsch parameter of a balanced unitary Fermi gas calculated according to Eq.~\eqref{eq:xi2} (filled circles), using the energies for $N$ trapped fermions reported in Refs.~\cite{Busch1998,Yin2015}. The error bars resulting from uncertainty in the few-body energies are smaller than the symbol sizes. The darker circles indicate closed shells of the corresponding non-interacting problem in the harmonic oscillator.
The diamond shows the experimental data point~\cite{Ku2012}, which has an error bar smaller than the symbol size.}
\end{figure}

The energy of a few harmonically trapped fermions in the unitary limit
has been accurately predicted in a series of works, starting with the
exact solution by Busch {\em et al.} for the two-spin $\up\down$
problem \cite{Busch1998}. Here, it was shown that the energy at
unitarity is exactly $2\omega$; that is, interactions have reduced the
energy from its value of $3\omega$ in the limit of weak attractive
interactions, $1/a \to -\infty$.  This reduction of the total energy
upon approaching unitarity is a generic feature, which is reflected in
the Bertsch parameter of the unpolarized many-body system, and in the
polaron energy, as we discuss in this section.

In the following, we focus on data for the ground-state energy from
precise few-body calculations
\cite{Busch1998,Werner2006,Daily2010,BlumeDaily2010,Blume2011,Rakshit2012,Bradly2014,Yin2015}.
This presents an alternative angle to more intensive numerical studies
of many trapped particles in the spin-balanced case
\cite{Blume2007,Bulgac2007,Jauregui2007,Chang2007,vonStecher2008,Zubarev2009,Nicholson2010,Carlson2014}
and for large spin polarizations \cite{Blume2008}.

In the limit $N \gg 1$, we know from LDA that the Bertsch parameter
relates the energy of the unitary Fermi gas to that of the
non-interacting system through $\sqrt{\xi}=\mathcal{E}/\mathcal{E}_0$
--- see Eq.~\eqref{eq:xi1}.  In order to minimize shell effects for
smaller $N$, it is convenient to think of the Bertsch parameter as
arising from an interaction energy shift compared with the
non-interacting limit $1/a \to - \infty$.  That is, we measure the
energy at unitarity with respect to the {\em exact} non-interacting
energy $\mathcal{E}_{\rm{NI}}$ of the few-body system. We then
normalize this shift by the non-interacting energy \eqref{eq:nonint}
calculated within LDA, as this does not contain any shell effects. In
this manner, we modify Eq.~\eqref{eq:xi1} to give
\begin{align}
    \xi=\left[1+(\mathcal{E}-\mathcal{E}_{\rm{NI}})/\mathcal{E}_0\right]^2,
\label{eq:xi2} 
\end{align}
which, of course, converges to Eq.~\eqref{eq:xi1} when $N\to\infty$.

The result of our analysis is shown in Fig.~\ref{fig:bertsch}. We see
that the Bertsch parameter obtained within few-body calculations
yields a value that is remarkably close to (within 15\% of) the value
reported in experiment \cite{Ku2012}
\begin{align}
    \xi=0.376(4),
    \label{eq:xiexp}
\end{align}
and corroborated by QMC calculations \cite{Carlson2011}.  Even the
simplest result, given by the exact two-body solution, has a relative
error of only about $7\%$, further supporting the central idea that
the Bertsch parameter is primarily determined by two-body correlations
in the gas.

Let us now turn to the high polarization limit, where $N_\down=1$. In
this case, the interaction energy shift at unitarity can be related to
the polaron energy in uniform space.  Most notably, it has been
predicted both in 3D~\cite{Blume2008} and in 1D~\cite{Levinsen2015}
that the trapped density distribution of the impurity in the many-body
limit is almost unaffected by strong interactions.  In other words,
the impurity is always confined to the center of the trap, even in the
presence of a medium. Consequently, we expect that the polaron
interaction energy is proportional to the chemical potential of the
majority fermions at the center of the trap, i.e.,
\begin{align}
  E_{\rm{pol}}=\mathcal{E}-\mathcal{E}_{\rm{NI}}=-A\varepsilon_F,
\end{align}
where we again use $\mathcal{E}_{\rm{NI}}$ as the exact energy of the
non-interacting system.  The coefficient of proportionality $A$ should
asymptote towards the same value as in the uniform system,
$E_{\rm{pol}}=-0.615E_F$ \cite{Vlietinck2013}.

\begin{figure}[t]
\centering
\includegraphics[width=\columnwidth]{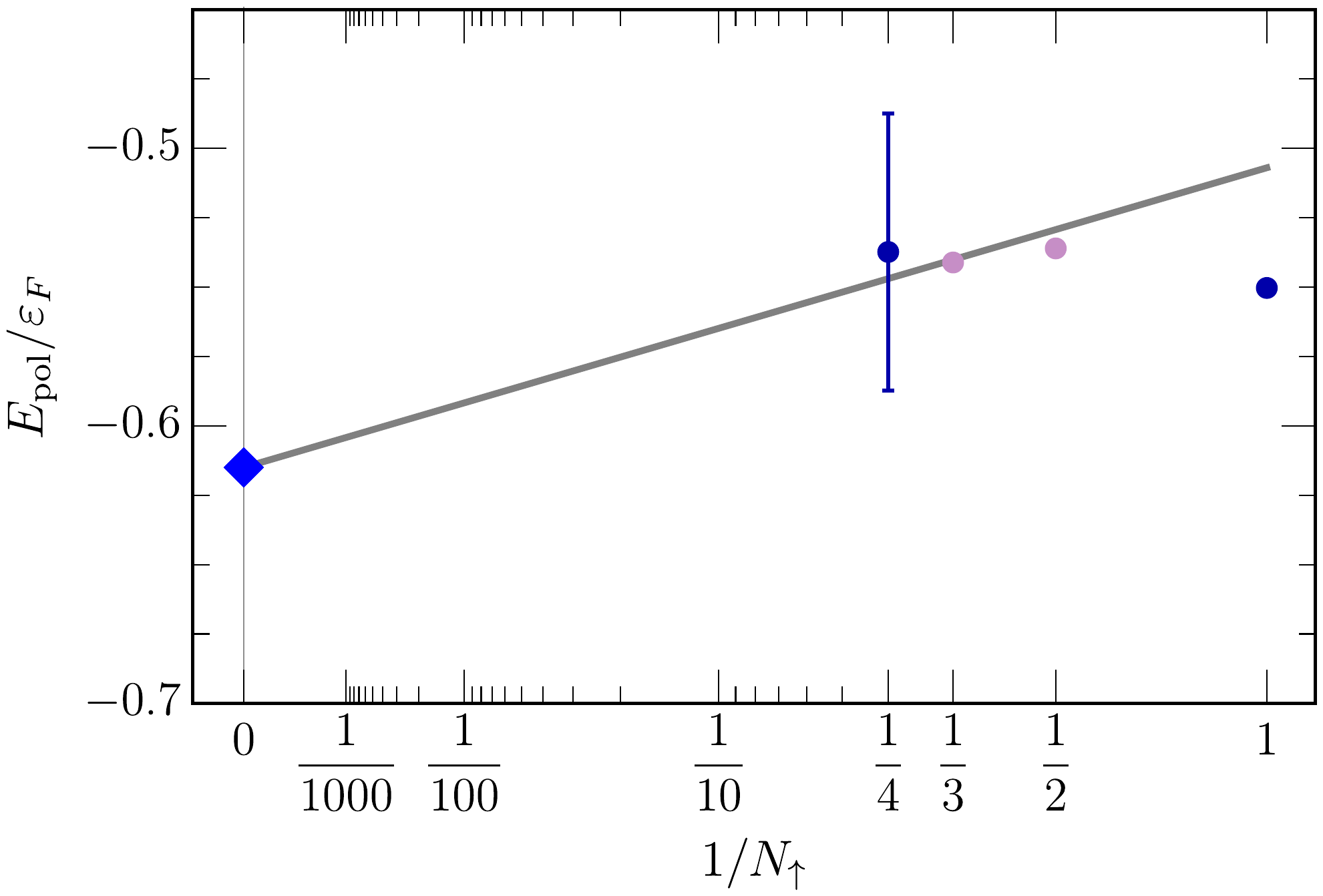}
\caption{\label{fig:epol} Polaron energy (circles) in units of the
  local spin-$\up$ Fermi energy at the center of the trap as a
  function of particle number $N_\up$. The darker circles indicate
  closed shells.  In the many-body limit of $N_\up\to\infty$, the
  polaron energy is expected to approach the uniform system result,
  $E_{\rm{pol}}=-0.615E_F$ \cite{Vlietinck2013}, indicated as a
  diamond. The few-body results are taken from
  Refs.~\cite{Busch1998,Werner2006,BlumeDaily2010,Rakshit2012}. The
  straight line is a fit to the data points for 2--4 majority
  particles using the form Eq.~\eqref{eq:fit}, keeping the known
  coefficient in the many-body limit fixed to $A=0.615$.}
\end{figure}

In Fig.~\ref{fig:epol}, we show the polaron energy obtained in this
manner from few-body calculations of the energy. The figure explicitly
shows that only very few spin-$\up$ particles are needed to obtain an
approximate value of the polaron energy. In particular, from the exact
solution of the two-body problem, we find
$E_{\rm pol}=-6^{-1/3}\varepsilon_F\approx-0.55\varepsilon_F$ which
has a relative error of only about $10\%$ compared with the
$N_\up\to\infty$ limit.  We note that a related analysis was carried
out by D.~Blume \cite{Blume2008}.

The few-body results for the ground-state energy also allow us to
estimate the polaron effective mass \cite{Lobo2006}. Within LDA, the
effective trapping potential experienced by the polaron is
approximately \cite{Recati2008}
\begin{align*}
V_{\rm{eff}}(\r)=\frac12m\omega^2r^2-A\mu(\r)=\frac12m(1+A)\omega^2r^2-A\varepsilon_F,
\end{align*}
and thus the ground-state energy of a particle of mass $m^*$ takes the
form $\frac32\sqrt{\frac{m}{m^*}(1+A)}\omega-A\varepsilon_F$. Using
the relation between the Fermi energy and the particle number in
Eq.~\eqref{eq:ktoN}, we thus arrive at
\begin{align} \label{eq:fit}
\frac{E_{\rm{pol}}}{\varepsilon_F}=-A+\frac32\sqrt{\frac{m}{m^*}(1+A)}(6N_\up)^{-1/3}.
\end{align}
This demonstrates that the leading order correction to the
$N_\up \to \infty$ limit for the polaron energy goes like
$N_\up^{-1/3}$ in the trapped system, as illustrated in
Fig.~\ref{fig:epol}. By fitting Eq.~\eqref{eq:fit} to the data (see
Fig.~\ref{fig:epol}), we obtain an effective polaron mass of
$m^*=1.26m$, which is in reasonable agreement with the result of QMC
calculations, where $m^*=1.197m$ \cite{Vlietinck2013}.

\section{Contact}
\label{sec:contact}

A quantity that is naturally associated with two-body physics is the
contact \cite{Tan2008,Tan2008b}.  As defined in Eq.~\eqref{eq:C}, it
is intrinsically a thermodynamic quantity.  However, it can also be
related to the probability of finding two particles at a separation
much less than the average interparticle spacing,
$C=16\pi^2\lim_{\r\to\0}r^2\left<n_\down(\r/2)n_\up(\r/2)\right>$,
with $n_\sigma(\r)$ the density of spin $\sigma$ particles at position
$\r$ \cite{Braaten2012}. This expression exposes the two-body nature
of the contact, even in a medium, and we now investigate how few-body
results for the contact compare with the many-body limit.

In the harmonically trapped system at unitarity, we may again obtain
the contact analytically for the two-body problem from the exact
solution \cite{Busch1998}, yielding
${\cal C}=4\pi\sqrt{2/\pi}\aho^{-1}$. For larger spin-balanced gases,
the trap averaged contact has been computed in few-body calculations
in Ref.~\cite{Yin2015}. These results may be related to the contact in
uniform space by applying the LDA --- see Eq.~\eqref{eq:Crelation}.

\begin{figure}[t]
\centering
\includegraphics[width=\columnwidth]{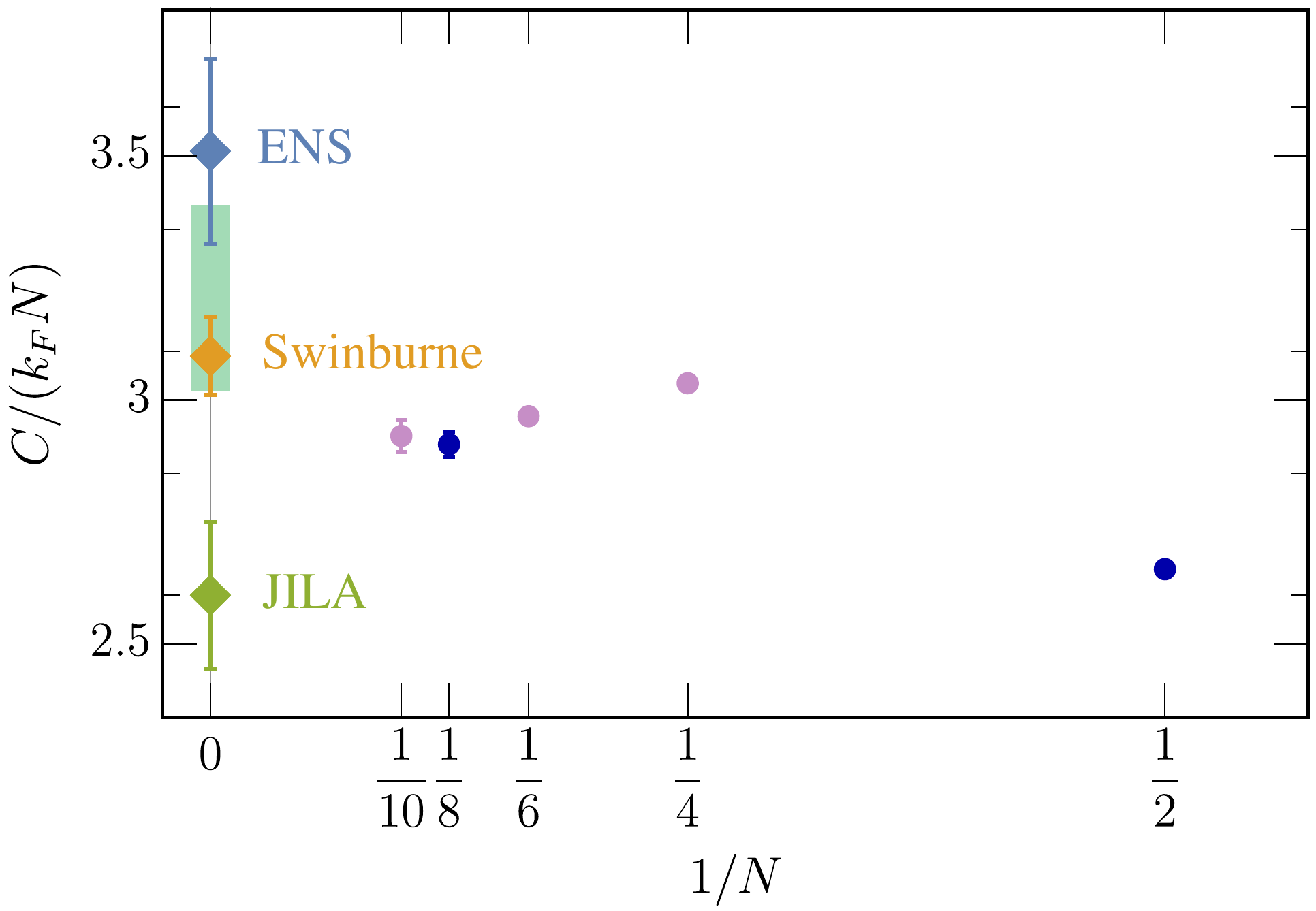}
\caption{\label{fig:cbal} Contact of the unpolarized unitary Fermi gas
  with $N$ particles. The few-body results \cite{Busch1998,Yin2015}
  are indicated by circles, where the darker circles denote closed
  shells. The experimental points (diamonds) are from
  Refs.~\cite{Navon2010,Sagi2012,Hoinka2013}. The light green shading
  indicates the region of various recent theoretical results
  \cite{Combescot2006,Haussmann2009,Gandolfi2011,Palestini2010,Hu2011,Hoinka2013}.}
\end{figure}

In Fig.~\ref{fig:cbal}, we show the resulting contact converted to the
uniform system. We see that the few-body results compare well with
recent low-temperature measurements: Experiments at the ENS found
$C/(N k_F)=3.51\pm0.19$ \cite{Navon2010,Braaten2012}, while the
Swinburne group measured $C/(N k_F)=3.09\pm0.08$ \cite{Hoinka2013}.
At a somewhat higher temperature ($T\sim0.16T_F$), the JILA group
found $C/(N k_F)\approx2.6\pm0.15$ \cite{Sagi2012}. We also note that
numerous authors have evaluated the contact theoretically, finding
values of $C/(N k_F)$ ranging from 3.01 to 3.40
\cite{Combescot2006,Haussmann2009,Gandolfi2011,Palestini2010,Hu2011,Hoinka2013}
(see, e.g., Ref.~\cite{Hu2011} for an extensive review). These values
are all compatible with the few-body results.

From our results for the contact, we can also estimate the value of
the pairing gap using the expression $C/V \approx 2m^2 \Delta^2$ from
Sec.~\ref{sec:unpol}.  Taking $C/(k_FN) \approx 3$ in the many-body
limit, we obtain $\Delta/E_F \approx \sqrt{2/\pi^2} \approx 0.45$,
which compares remarkably well with the experimentally determined
value of $\Delta/E_F \approx 0.44$ \cite{Schirotzek2008}. While this
is certainly not a formal proof, it does raise the question of whether
the contact at unitarity is intimately connected to the pairing gap.

\begin{figure}[t]
\centering
\includegraphics[width=\columnwidth]{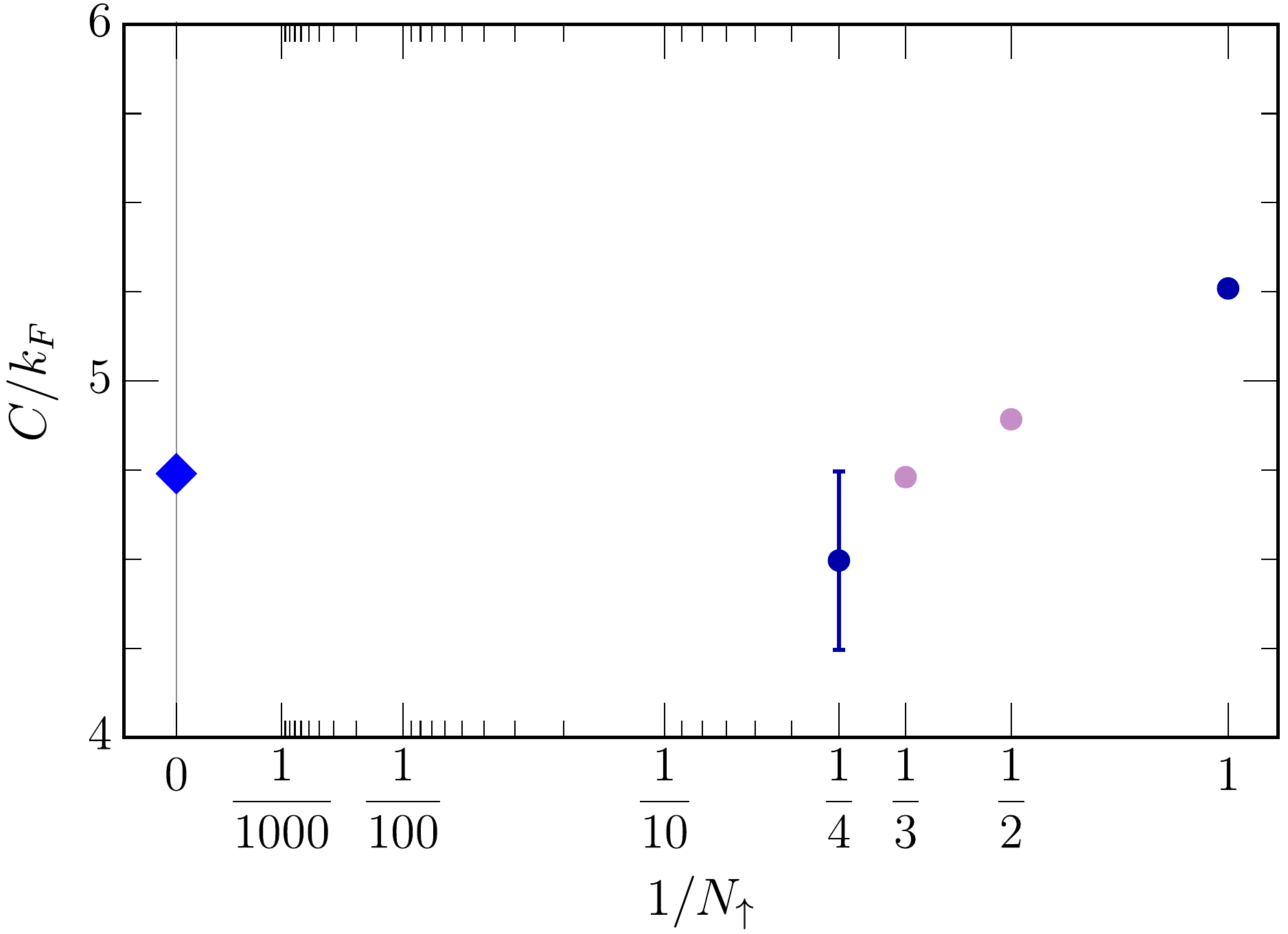}
\caption{\label{fig:cpol} Contact of a single $\down$ impurity in a
  gas with $N_\up$ majority fermions. The few-body results
  \cite{Busch1998,Werner2006,Yin2015,Yan2013,BlumePC} are marked with
  circles, the darker circles corresponding to closed shells. The
  diamond indicates the thermodynamic result in uniform space obtained
  from the 1PH Ansatz \cite{Punk2009}.  }
\end{figure}

Turning now to the polaron limit, we again take advantage of how the
impurity is primarily located at the center of the trap. Therefore, in
this case, we derive the uniform contact by simply equating the local
Fermi momentum $\kappa_F$ at the center of the trap with its uniform
space counterpart, $k_F$.  Moreover, from Eq.~\eqref{eq:fit}, we see
that the leading order correction to the many-body limit once again
scales as $N_\up^{-1/3}$ in the trapped system.  In
Fig.~\ref{fig:cpol} we show the results of few-body calculations for
the trapped contact \cite{Yan2013}, where we have converted the
results using the relation between $\kappa_F$ and the majority
particle number, Eq.~\eqref{eq:ktoN}.  We see that the few-body
contact takes values ranging between $4\kappa_F$ and $6\kappa_F$.  In
this case, the contact has not been measured experimentally, however,
it has been calculated within the ansatz described in Section
\ref{sec:chevy} to be $C=4.74 k_F$~\cite{Punk2009}.  Given that this
ansatz produces very reliable results for the other quasiparticle
parameters, there is no a priori reason to believe that this result is
inaccurate. The agreement with the results from the harmonically
trapped few-body calculation is also quite reasonable.

\section{Wave-function overlaps}
\label{sec:residue}

The squared wave-function overlap between a strongly interacting state
and its non-interacting counterpart is an important quantity in
many-body physics. For instance, in the case of the polaron, this
overlap gives its quasiparticle weight, or residue.  The overlap can
be measured in Rabi oscillations as a shift of the oscillation
frequency \cite{Kohstall2012,Parish2016,Scazza2016}.  While the
polaron residue of the $N_{\uparrow}$+1 system (a system of $N_\up$
spin $\up$ fermions and one spin $\down$ impurity) in the many-body
limit $N_{\uparrow}\rightarrow \infty$ has been actively studied both
experimentally~\cite{Schirotzek2009,Kohstall2012,Scazza2016} and
theoretically~\cite{Punk2009}, it is still an open question how the
polaron residue can be understood from a few-body perspective.

For two-particle systems, the ``residue'' of the ground-state wave
function at unitarity is obtained analytically as
\begin{align}
\label{eq:Z2b}Z=\frac{3}{4} \  \ (\textrm{uniform}); \hspace{5mm}
\mathcal{Z}=\frac{2}{\pi}\simeq0.64 \  \ (\mathrm{trap}).
\end{align}
The uniform system result can be calculated by noting that the
relative wave function of two non-interacting particles at zero energy
is a constant, while for two particles at unitarity it is proportional
to $1/r$. In a large sphere of volume $V=4\pi L^3/3$, we then find The
trapped result in Eq.~\eqref{eq:Z2b} can be similarly obtained from
the analytic solution of the two-body problem in a harmonic
trap~\cite{Busch1998}. However, such exact calculations will be
exponentially more challenging as the number of particles increases.

In this section, we use a hyper-spherical coordinates method to study
the squared overlap between the ground state wave function of the
unitary Fermi gas and its non-interacting counterpart in a harmonic
trap as well as in a uniform system. We present a general formalism to
calculate it for arbitrary particle numbers based on the separability
of the hyper-radial and hyper-angular parts of the wave functions in
the unitary and non-interacting systems. We derive a rigorous upper
bound of the squared overlap, $Z_{>} \ge Z$, as a function of the
energy of the trapped unitary and non-interacting systems. We evaluate
this for systems with $N_{\uparrow},N_{\downarrow}\lesssim 5$ using
energies of the few-body unitary Fermi gas in a harmonic
trap~\cite{Rakshit2012,Yin2015}, and then extend the calculation to
$N\gg1$ using LDA.  We find that our rigorous upper bound for the
polaron system approaches $Z_{>}\rightarrow 1$ as
$N_{\uparrow}\rightarrow \infty$ for both trapped and uniform systems
and thus does not actually provide a bound. On the other hand, the
upper bound decreases exponentially, approaching
$\mathcal{Z}_{>}\rightarrow 0$ for a trapped balanced system,
dictating that the squared overlap must vanish in the many-body limit
in the trap. Conversely, we find that it approaches a finite value
$Z_{>}\rightarrow 0.942...$ in the uniform system.

\subsection{Hyper-spherical method to evaluate the squared overlap of the wave functions}
\subsubsection{Harmonically trapped systems}
For any $N_{\downarrow}+N_{\uparrow}$ system in an isotropic harmonic
trap with unitary zero-range interactions, the wave function of $N$
particles at positions $\r_i$, $i=1,\cdots,N$, separates into
center-of-mass, hyper-radial, and hyper-angular
motions~\cite{Castin2012}:
\begin{equation}
\label{eq:TotWF}\Psi = \psi_{\mathrm{CM}}(\bm{R}_{\mathrm{CM}})\frac{\mathcal{F}_s(R)}{R^{\frac{3N-4}{2}}}\Phi_s(\Omega).
\end{equation}
Here, $\bm{R}_{\mathrm{CM}}=\frac{1}{N}\sum_{i=1}^{N}\bm{r_i}$ is the
center-of-mass coordinate,
$ R^2=\sum_{i=1}^{N}(\bm{r_i}-\bm{R}_{\mathrm{CM}})^2$ is the
(squared) hyper radius describing the overall size of the $N$-body
system, and $\Omega$ denotes a set of $(3N-4)$-dimensional
hyperangles, describing rotation and deformation degrees of
freedom. The hyper-radial equation takes the form
\begin{align}
\left[-\frac{1}{2m}\frac{d^2}{d R^2} \!+\! \frac{s^2 -\frac{1}{4}}{2m R^2}\!+\!\frac{m\omega^2 R^2}{2}\right]\!\mathcal{F}_s(R)\! =\! (\mathcal{E}\!-\!\mathcal{E}_{\mathrm{CM}})\mathcal{F}_s(R),
\label{eq:hyperR}
\end{align}
where $\mathcal{E}_{\mathrm{CM}}$ is the energy of the center-of-mass
motion, and $s^2$ is the hyper-angular eigenvalue obtained by solving
the hyper-angular equation. The separability of the hyper-radial and
hyper-angular equations originates from the SO(2,1) dynamical symmetry
associated with the scale invariance of the unitary system in an
isotropic harmonic trap~\cite{Castin2012}, and it does not generally
hold for other confining potentials, e.g., an anisotropic trap or a
box-shaped trap~\cite{PhysRevLett.110.200406}.

For the equal-mass unitary two-component Fermi system, it is believed
that the hyper-angular eigenvalue is positive $s^2>0$, and the Efimov
effect~\cite{Efimov1970a,braaten2006universality,ferlaino2010trend,naidon2016efimov,Efimov1973,PhysRevA.67.010703,Castin2010}
does not occur for any number of particles. Though this has not been
generally proved, it has been shown numerically for
$N_{\uparrow},N_{\downarrow}\lesssim 5$~\cite{Rakshit2012,Yin2015} and
mathematically for the $N_{\uparrow}$+1 system with any $N_{\uparrow}$
~\cite{moser2016stability}. Regular solutions of the hyper-radial
equation for $s^2>0$ are obtained as~\cite{Werner2006}:
\begin{equation}
\label{eq:hypradwf_trap}    \mathcal{F}_{s}(R)=R^{s+\frac{1}{2}} e^{-\frac{R^2}{2a_{\rm ho}^2} }L_{q}^{(s)}\left(\frac{R^2}{a_{\rm ho}^2} \right),
\end{equation}
where $L_{q}^{(s)}$ is the generalized Laguerre polynomial, and
$q=0,1,2....$ is a quantum number characterizing the SO(2,1) ladder of
the trapped energy spectrum:
$\mathcal{E}-\mathcal{E}_{\mathrm{CM}}=(s+2q+1) \omega$.  We focus on
the ground state $q=0$, which simplifies the above wave function since
$L_{q=0}^{(s)}=1$.

We note that the wave function and energy of the non-interacting
system have the same form as those of the unitary system in
Eqs.~(\ref{eq:TotWF},\ref{eq:hypradwf_trap}). The only difference is
the change in the hyper-angular eigenvalue $s\rightarrow s_0$, which
is related to the non-interacting energy $\mathcal{E}_{\rm{NI}}$ via
$\mathcal{E}_{\rm{NI}}-\mathcal{E}_{\mathrm{CM}}=(s_0+2q+1) \omega$.

Using separability and the similarity of the wave functions of the
unitary and non-interacting systems in the trap, and
rewriting 
$
\int \prod_{i=1}^{N} d^3\bm{r}_i =\int d^3 \bm{R}_{\mathrm{CM}} R^{3N-4}dR\, d\Omega$, the squared wave-function overlap between unitary and non-interacting ground states reads
\begin{equation}
  \mathcal{Z}=
  \frac{
    \mathcal{I}^2_{s,s_0}
  }{\mathcal{I}_{s,s}\mathcal{I}_{s_0,s_0}} \frac{|\langle \Phi_{s_0}|\Phi_{s}\rangle |^2}{\langle \Phi_{s_0}|\Phi_{s_0}\rangle \langle \Phi_{s}|\Phi_{s}\rangle}.
 \label{eq:olaps}
\end{equation}
Here
\begin{equation}
\mathcal{I}_{s_1,s_2}=\int d R\, \mathcal{F}_{s_1}(R)\mathcal{F}_{s_2}(R) = \frac{\aho^{s_1+s_2+2}}{2}\Gamma \left(\tfrac{s_1+s_2}{2}+1 \right),
\end{equation}
$\langle \Phi_{s_1}|\Phi_{s_2}\rangle = \int d\Omega\,
\Phi_{s_1}^*(\Omega) \Phi_{s_2}(\Omega)$,
and we have assumed that the non-interacting and unitary systems are
in the same center-of-mass motion state. We note that
Eq.~\eqref{eq:olaps} holds rigorously for any number of particles,
few-body or many-body, and any population imbalance for equal-mass
two-component Fermi systems. Since the derivation relies only on the
separability of the wave function, a similar result can be obtained,
with properly defined mass-scaled Jacobi coordinates and
hyper-radius~\cite{braaten2006universality,naidon2016efimov}, for any
other scale-invariant few-body or many-body systems
\footnote{Equation~\eqref{eq:olaps} remains valid even when the
  $N$-body Efimov
  effect~\cite{Efimov1970a,braaten2006universality,ferlaino2010trend,naidon2016efimov}
  occurs, as long as any $n<N$ subsystem does not show the Efimov
  effect. As examples, it is valid for a system of 3 identical
  bosons~\cite{Efimov1970a}, since the 3-body parameter appears only
  in the hyper-radial wave function. For a mass-imbalanced 2+1 Fermi
  system, in which the Efimov effect occurs for
  $m_{\uparrow}/m_{\downarrow}>13.606...$~\cite{Efimov1973,PhysRevA.67.010703},
  it holds for the entire mass ratio. We note however that when a
  $n<N$ subsystem shows the Efimov effect, the hyper-angular wave
  function depends on an additional $n$-body parameter, and
  Eq.~\eqref{eq:olaps} breaks down. A prime example would be the
  mass-imbalanced 3+1 Fermi system for
  $m_{\uparrow}/m_{\downarrow}>13.606....$, where the three-body
  parameter non-trivially appears in the hyper-radial and
  hyper-angular wave functions, so that the total wave function cannot
  be written as in the separable form of Eq.~\eqref{eq:TotWF}. For
  $m_{\uparrow}/m_{\downarrow}<13.606....$, on the other hand,
  Eq.~\eqref{eq:olaps} is valid, even when the four-body Efimov states
  appear~\cite{Castin2010}.}

From Eq.~\eqref{eq:olaps} we can derive a rigorous upper bound, i.e.,
$\mathcal{Z}_{>} \ge \mathcal{Z}$, for the squared overlap of the wave
functions \footnote{An upper bound similar to
  Eq.~\eqref{eq:Zumit_trap} can be derived even in the presence of an
  $N$-body Efimov effect ($s^2<0$) as long as the Efimov effect does
  not occur in any subsystem (see the footnote above), by replacing
  the hyper-radial wave function in Eq.~\eqref{eq:hypradwf_trap} by
  the corresponding hyper-radial solution
  $\mathcal{F}_{s}(R)=R^{-\frac{1}{2}}W_{\frac{(\mathcal{E}-\mathcal{E}_{\mathrm{CM}})}{2
      \omega},\frac{s}{2}}\left(\frac{R^2}{a_{\rm ho}^2}
  \right)$~\cite{Werner2006},
  supplemented with an $N$-body boundary condition at small
  hyper-radius. Here, $W$ is the Whittaker function.}:
\begin{equation}
\label{eq:Zumit_trap}\mathcal{Z}_{>}
=\frac{\left[\Gamma \left(\frac{s_0+s}{2}+1 \right)\right]^2}{\Gamma \left(s+1 \right)\Gamma \left(s_0+1 \right)}.
\end{equation}
Physically, $\mathcal{Z}_{>}$ represents the overall spatial overlap
of the hyper-radial part of the two wave functions. Indeed, if the
energy difference between the non-interacting and unitary systems is
large, their sizes, characterized by the
$\displaystyle R^{s+\frac{1}{2}}$ term in
Eq.~\eqref{eq:hypradwf_trap}, become substantially different,
rendering $\mathcal{Z}_>$ small. We can compute the upper bound
$\mathcal{Z}_{>}$ once we know the energy eigenvalues for the unitary
and non-interacting harmonically trapped system. However, note that
it totally dismisses few- and many-body correlations appearing in the
hyper-angular wave functions.

\subsubsection{Uniform systems}

Interestingly, the hyper-angular equation is the same for the
harmonically trapped system and the uniform system, so that they
possess the same hyper-angular functions and eigenvalues as those of
the trapped systems~\cite{Castin2012}. The hyper-radial and
hyper-angular parts are also separable.  The only difference between
the trapped and uniform systems is that the hyper-radial part $F_s(R)$
now satisfies Eq.~\eqref{eq:hyperR} at $\omega=0$.  The hyper-radial
wave function which is regular at $R\rightarrow 0$ is then
$F_s(R)\propto \sqrt{R} J_{s}(kR)$ where $J_n$ is the Bessel function,
and $\displaystyle k^2 = 2m(E-E_{\mathrm{CM}})$. At zero energy, in
particular, it becomes $F_s(R)=R^{s+\frac{1}{2}}$.

Similarly to the trapped system, the squared wave-function overlap for
the uniform system at zero energy is
\begin{align}
\label{eq:Zhom_intcalc1}Z&= \frac{I_{s,s_0}^2}{I_{s,s}I_{s_0,s_0}}\frac{|\langle \Phi_{s_0}|\Phi_{s}\rangle |^2}{\langle \Phi_{s_0}|\Phi_{s_0}\rangle \langle \Phi_{s}|\Phi_{s}\rangle},
\end{align}
where
\begin{equation}
I_{s_1,s_2}=\int d R\, F_{s_1}(R)F_{s_2}(R)=\int d R \,R^{s_1+s_2+1}.
\end{equation}
This integral is formally divergent at large $R$ but this is
compensated by the same divergence appearing in the denominator in
Eq.~\eqref{eq:Zhom_intcalc1}. We thus relate the squared overlap of
the unitary and non-interacting ground state wave functions in the
uniform ($Z$) and trapped ($\mathcal{Z}$) systems:
\begin{equation}
\label{eq:Zrelation_TrapHom}Z=\frac{\Gamma \left(s+2 \right)\Gamma \left(s_0+2 \right)}{\left[\Gamma \left(\frac{s_0+s}{2}+2 \right)\right]^2} \mathcal{Z}.
\end{equation}
Importantly, this exact relation enables us to extract the uniform
space result from only the energy (corresponding to $s$) and the
harmonic trap overlap. We emphasize that the above arguments and
results are valid for any number of particles, for both few- and
many-body systems, in marked contrast to LDA, which is valid only for
systems with large number of particles.

From Eq~\eqref{eq:Zrelation_TrapHom}, one also finds the upper limit
of $Z$ for the uniform system, $\displaystyle Z_{>} \ge Z$, where
\begin{equation}
  \label{eq:Zumit_hom}Z_{>}= \frac{4(s+1)(s_0+1)}{(s+s_0+2)^2} = \frac{\Gamma \left(s+2 \right)\Gamma \left(s_0+2 \right)}{\left[\Gamma \left(\frac{s_0+s}{2}+2 \right)\right]^2} \mathcal{Z}_{>}.
\end{equation}

\subsection{Applications to few- and many-body systems}

\begin{figure}[t]
\centering
\includegraphics[width=\columnwidth]{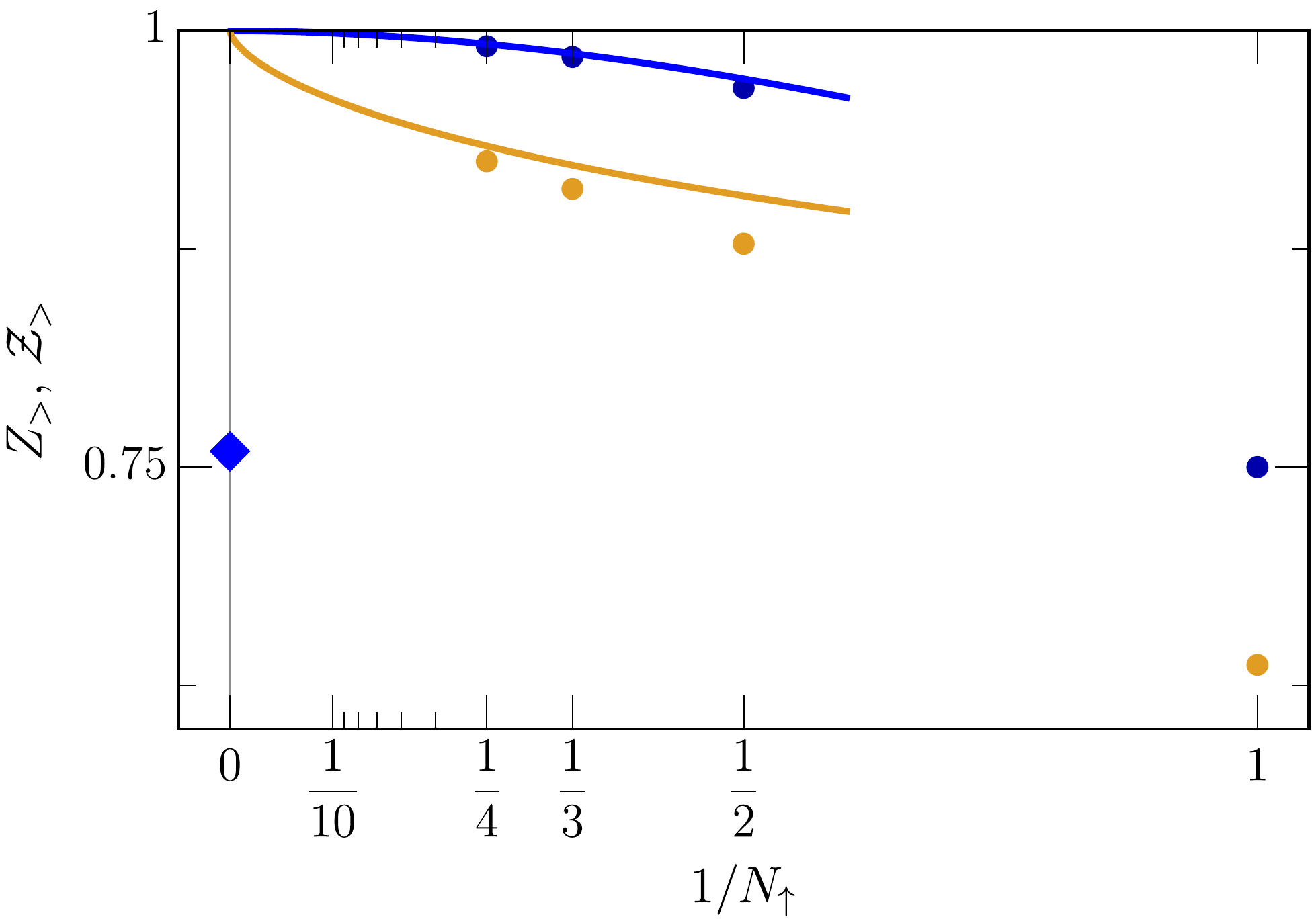}
\caption{\label{fig:Zpol} Upper limit of the polaron residue for the
  trapped (yellow) and uniform systems (blue), as a function of the
  number of majority particles $N_\up$. The circles represent the
  few-body results in Eqs.~(\ref{eq:Zumit_trap},\ref{eq:Zumit_hom})
  evaluated with the energies calculated in
  Refs.~\cite{BlumeDaily2010,Rakshit2012,Yin2015}. The error bars are
  smaller than the symbol sizes. The solid curves are the many-body
  result in Eqs.~(\ref{eq:Zpol_umit_trap},\ref{eq:Zpol_umit_hom}). For
  $N_\up\to\infty$, the polaron residue in the uniform system
  $Z_{\rm{pol}}=0.759$~\cite{Vlietinck2013} is indicated as a blue
  diamond. For the upper bound curve we have used the constants
  $m^*/m =1.198$ and
  $A = 0.615$~\cite{Prokofev2008,Combescot2008,Vlietinck2013}.}
\end{figure}

\subsubsection{Polaron residue in the limit of large spin imbalance}
We now apply our hyperspherical approach to evaluate the upper bound
for the polaron residue at unitarity. To this end, we use the energy
of the trapped $N_{\uparrow}+1$ Fermi system at large $N_{\uparrow}$:
$\mathcal{E}=\mathcal{E}_{0\up}-A \mu +\frac{3}{2}\omega
\sqrt{\frac{m}{m^*}(1+A)}$
--- see the discussion in Sec.~\ref{sec:energy}.  Here we take into
account the correction to the LDA~\cite{Blume2012} for both the
non-interacting energy,
$\mathcal{E}_{0\up} = \frac{\omega}{8}(6N_{\uparrow})^{4/3}\left[1 +
  \frac{1}{2} (6 N_{\uparrow})^{-2/3}\right]$,
and the chemical potential
$\mu=\partial \mathcal{E}_{0\up}/\partial N_{\uparrow}$.  From the
polaron energy, or rather from
$(s_0+1)\omega=\mathcal{E}_{0\up}-\frac{3}{2}\omega$ and
$(s+1)\omega=\mathcal{E}-\frac{3}{2}\omega $, one can evaluate the
upper bound of its residue.  Using Eq.~\eqref{eq:Zumit_trap} together
with the Stirling formula for the Gamma function, we obtain
\begin{equation}
  \label{eq:Zpol_umit_trap}\mathcal{Z}_{>} = 1 - \frac{2^{\frac{1}{3}}A^2}{(3N_{\uparrow})^{\frac{2}{3}}}  + \frac{A}{N_{\uparrow}}\left[\sqrt{\tfrac{m}{m^*}(1 + A)} -1\right]+O\left(N_{\uparrow}^{-\frac{4}{3}}\right)
\end{equation}
and
\begin{equation}
  \label{eq:Zpol_umit_hom}Z_{>}=1 - \frac{4 A^2}{9 N_{\uparrow}^2} + \frac{2^{\frac{5}{3}} A }{3^{\frac{4}{3}} N_{\uparrow}^{\frac{7}{3}}}\left[\sqrt{\tfrac{m}{m^*}(1 + A)}-1\right]+O\left(N_{\uparrow}^{-\frac{8}{3}}\right)
\end{equation}
for the trapped and uniform systems, respectively.

In Fig.~\ref{fig:Zpol} we show these $N_\up \gg 1$ upper bounds for
the polaron (solid lines), together with few-body results (circles)
obtained from Eqs.~(\ref{eq:Zumit_trap},\ref{eq:Zumit_hom}) and
few-body data.  We see that these calculations match quite well for
intermediate $N_\up$.  We note that for two-body systems (right-most
circles), our upper bounds are in fact equal to the exact values in
Eq.~\eqref{eq:Z2b}. Quite remarkably, the exact residue of $3/4$ from
the two-body problem in uniform space is very close to the expected
many-body limit of $Z=0.759$ \cite{Vlietinck2013}.  It will be of
interest to determine whether this close agreement between the few-
and many-body limits holds for intermediate particle numbers. From
Eq.~\eqref{eq:Zrelation_TrapHom}, we further conclude that the polaron
residues in both the uniform and the trapped systems approach the same
value in the many-body limit, which is consistent with the expectation
that the impurity is primarily located at the center of the trap.

\subsubsection{Unpolarized many-body system}

The energy of the trapped unitary system is characterized by the
Bertsch parameter via $\mathcal{E} = 2\sqrt{\xi}\mathcal{E}_{0\up}$,
from which one finds
$(s+1)\omega = 2\sqrt{\xi}\mathcal{E}_{0\up}-\frac{3}{2}\omega $ and
$(s_0+1)\omega = 2\mathcal{E}_{0\up}-\frac{3}{2}\omega$.  Substituting
these values into Eq.~\eqref{eq:Zumit_trap}, we find the upper bound
in the trap
\begin{align}
\label{eq:Zeqp_umit_trap}\mathcal{Z}_{>} =\exp\left[ -d_{\xi} N^{4/3}\left(1 + \frac{(3N)^{-2/3}}{2}\right)+ O\left(1\right)\right]
\end{align}
where
\begin{align*}
d_{\xi}&= \frac{3^{\frac{4}{3}}}{4}\left[ (1+\sqrt{\xi})\log \left(\tfrac{2}{1+\sqrt{\xi}}\right) +  \sqrt{\xi} \log  \sqrt{\xi} \right] \simeq 0.0506.
\end{align*}
and we have used the Bertsch parameter in Eq.~\eqref{eq:xiexp}.
Similarly, we find the upper bound for a uniform system
\begin{equation}
\label{eq:Zeqp_umit_hom}Z_{>}=\frac{4\sqrt{\xi}}{(1+\sqrt{\xi})^2}+O(N^{-\frac{4}{3}})\simeq 0.942+O(N^{-\frac{4}{3}}).
\end{equation}

\begin{figure}[t]
\centering
\includegraphics[width=\columnwidth]{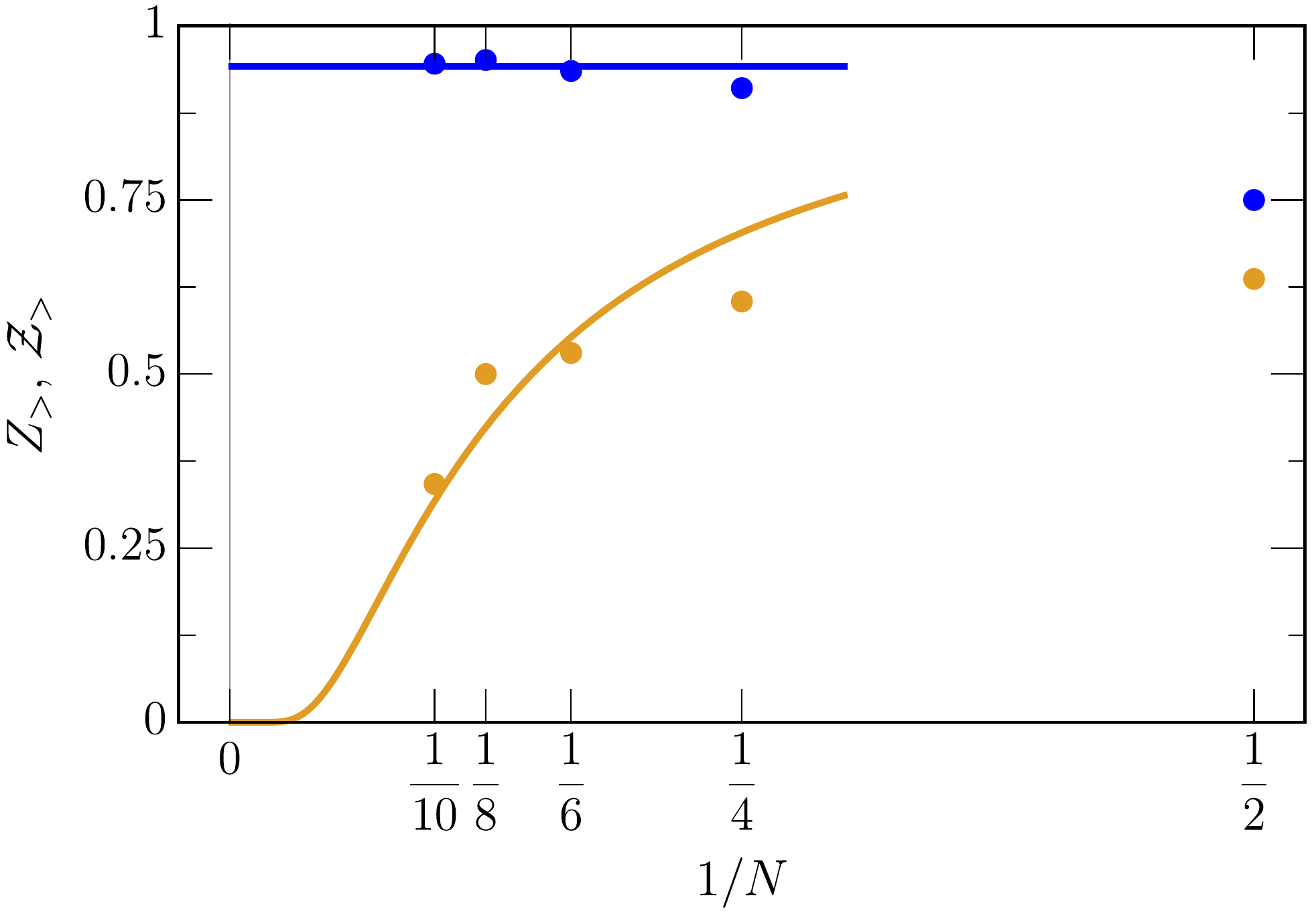}
\caption{\label{fig:Zbal} Upper limit of the squared overlap between
  the interacting and non-interacting ground states for the
  spin-balanced system in a trap (yellow) and in uniform space
  (blue). The circles represent the few-body results in
  Eqs.~(\ref{eq:Zumit_trap},\ref{eq:Zumit_hom}) evaluated with the
  energies calculated in Refs.~\cite{Rakshit2012,Yin2015}, and the
  solid curves are the many-body results in
  Eqs.~(\ref{eq:Zeqp_umit_trap},\ref{eq:Zeqp_umit_hom}). The error
  bars resulting from uncertainty in the few-body energies are smaller
  than the symbol sizes.}
\end{figure}

The upper bounds for the many-body squared overlaps are shown in
Fig.~\ref{fig:Zbal} as solid lines, together with results from
few-body physics (circles). Once again, we see that the few- and
many-body physics agree well for intermediate particle numbers. We
emphasize that our results are strict upper bounds, and they thus
prove that the trapped ground state at unitarity has zero overlap with
its non-interacting counterpart in the many-body limit. On the other
hand, the upper bound $Z_>$ remains finite for the uniform system,
which implies that the hyper-angular component dominates the behaviour
of the overlap for large $N$ in this case.

\section{Conclusions and outlook}
\label{sec:conclusions}

In this topical review paper, we have revisited the ground-state
properties of the unitary Fermi gas from the perspective of few-body
physics.  From the energies of unpolarized trapped few-body systems,
we found that we could extract values for the Bertsch parameter that
all lie within 15\% of that obtained from precision
experiments~\cite{Ku2012}.  Indeed, the simplest two-body result only
deviated from the established value by 7\%, lending weight to the idea
that the Bertsch parameter is primarily determined by two-body
correlations.  In the limit of large spin imbalance, we observed a
similar accuracy for the polaron energy at unitarity, and we
determined its dependence on spin-$\up$ particle number for
$N_\up \gg 1$.  Furthermore, we could obtain a reasonable estimate for
the polaron's effective mass by fitting the energy as a function of
$N_\up$.

We also investigated the contact, another key thermodynamic quantity
that characterizes the unitary Fermi system. We found that the
few-body results for the unpolarized system converged particularly
rapidly, with the values for the contact lying well within the range
of results determined in the literature for the many-body system. We
could also extract a pairing gap from the contact that was consistent
with the experimentally determined value~\cite{Schirotzek2008}.  It is
thus conceivable that the pairing gap is intimately related to the
contact in the unitary ground state.

Finally, we analyzed the squared wave-function overlap between the
ground state of the unitary Fermi system and its non-interacting
counterpart.  Here, we showed that the overlap for the two-body system
in uniform space is remarkably close to the predicted polaron residue
in the many-body limit.  We also derived a rigorous upper bound of the
squared overlap for both the trapped and uniform systems.  Using this,
we proved that the trapped unpolarized ground state at unitarity has
zero overlap with the non-interacting state in the many-body limit.

The successful use of few-body systems to infer many-body properties
suggests that the unitary equation of state at zero temperature is
dominated by few-body correlations.  Thus, a natural question is
whether this generally holds for a range of systems or whether it is
specific to the unitary Fermi gas.  In particular, is this behavior
generic for any scattering length?  Are Fermi statistics required
and/or is it necessary to have interactions that are sufficiently
short ranged?  Could the harmonic trapping potential commonly used in
few-body calculations be particularly beneficial in minimizing
finite-size effects?  Finally, it would be interesting to investigate
whether there are any other observables at zero temperature (e.g.,
dynamical quantities) which \textit{cannot} be captured within the
few-body framework.

\section{Acknowledgements}
We are grateful to D.~Blume for sharing her few-body data with us. We
would also like to thank D.~Blume, C.~Bradly, and A.~Martin for
fruitful discussions.  JL, SE, and MMP acknowledge financial support
from the Australian Research Council via Discovery Project
No.~DP160102739.  PM acknowledges funding from a ``Ram\'on y Cajal"
fellowship, from MINECO (Severo Ochoa SEV-2015-0522, and FOQUS
FIS2013-46768), Generalitat de Catalunya (SGR 874 and CERCA), and the
Fundaci\'o Privada Cellex.  This research was supported in part by the
National Science Foundation under Grant No.~NSF PHY11-25915.  All the
authors wish to thank the KITP for the generous hospitality during the
program ``Universality in Few-Body Systems''.

\bibliography{unitaryFG}

\begin{thebibliography}{89}%
\makeatletter
\providecommand \@ifxundefined [1]{%
 \@ifx{#1\undefined}
}%
\providecommand \@ifnum [1]{%
 \ifnum #1\expandafter \@firstoftwo
 \else \expandafter \@secondoftwo
 \fi
}%
\providecommand \@ifx [1]{%
 \ifx #1\expandafter \@firstoftwo
 \else \expandafter \@secondoftwo
 \fi
}%
\providecommand \natexlab [1]{#1}%
\providecommand \enquote  [1]{``#1''}%
\providecommand \bibnamefont  [1]{#1}%
\providecommand \bibfnamefont [1]{#1}%
\providecommand \citenamefont [1]{#1}%
\providecommand \href@noop [0]{\@secondoftwo}%
\providecommand \href [0]{\begingroup \@sanitize@url \@href}%
\providecommand \@href[1]{\@@startlink{#1}\@@href}%
\providecommand \@@href[1]{\endgroup#1\@@endlink}%
\providecommand \@sanitize@url [0]{\catcode `\\12\catcode `\$12\catcode
  `\&12\catcode `\#12\catcode `\^12\catcode `\_12\catcode `\%12\relax}%
\providecommand \@@startlink[1]{}%
\providecommand \@@endlink[0]{}%
\providecommand \url  [0]{\begingroup\@sanitize@url \@url }%
\providecommand \@url [1]{\endgroup\@href {#1}{\urlprefix }}%
\providecommand \urlprefix  [0]{URL }%
\providecommand \Eprint [0]{\href }%
\providecommand \doibase [0]{http://dx.doi.org/}%
\providecommand \selectlanguage [0]{\@gobble}%
\providecommand \bibinfo  [0]{\@secondoftwo}%
\providecommand \bibfield  [0]{\@secondoftwo}%
\providecommand \translation [1]{[#1]}%
\providecommand \BibitemOpen [0]{}%
\providecommand \bibitemStop [0]{}%
\providecommand \bibitemNoStop [0]{.\EOS\space}%
\providecommand \EOS [0]{\spacefactor3000\relax}%
\providecommand \BibitemShut  [1]{\csname bibitem#1\endcsname}%
\let\auto@bib@innerbib\@empty
\bibitem [{\citenamefont {Giorgini}\ \emph {et~al.}(2008)\citenamefont
  {Giorgini}, \citenamefont {Pitaevskii},\ and\ \citenamefont
  {Stringari}}]{Giorgini2008RMP}%
  \BibitemOpen
  \bibfield  {author} {\bibinfo {author} {\bibfnamefont {S.}~\bibnamefont
  {Giorgini}}, \bibinfo {author} {\bibfnamefont {L.~P.}\ \bibnamefont
  {Pitaevskii}}, \ and\ \bibinfo {author} {\bibfnamefont {S.}~\bibnamefont
  {Stringari}},\ }\bibfield  {title} {\bibinfo {title} {\emph {Theory of
  ultracold atomic Fermi gases}},\ }\href {\doibase 10.1103/RevModPhys.80.1215}
  {\bibfield  {journal} {\bibinfo  {journal} {Rev. Mod. Phys.}\ }\textbf
  {\bibinfo {volume} {80}},\ \bibinfo {pages} {1215} (\bibinfo {year}
  {2008})}\BibitemShut {NoStop}%
\bibitem [{\citenamefont {Parish}(2015)}]{parish2015}%
  \BibitemOpen
  \bibfield  {author} {\bibinfo {author} {\bibfnamefont {M.~M.}\ \bibnamefont
  {Parish}},\ }\bibfield  {title} {\bibinfo {title} {\emph {The BCS--BEC
  crossover}},\ }in\ \href@noop {} {\emph {\bibinfo {booktitle} {Quantum Gas
  Experiments -- Exploring Many-Body States}}},\ \bibinfo {editor} {edited by\
  \bibinfo {editor} {\bibfnamefont {P.}~\bibnamefont {T\"orm\"a}}\ and\
  \bibinfo {editor} {\bibfnamefont {K.}~\bibnamefont {Sengstock}}}\ (\bibinfo
  {publisher} {Imperial College Press},\ \bibinfo {year} {2015})\BibitemShut
  {NoStop}%
\bibitem [{\citenamefont {Heiselberg}(2012)}]{Heiselberg2012}%
  \BibitemOpen
  \bibfield  {author} {\bibinfo {author} {\bibfnamefont {H.}~\bibnamefont
  {Heiselberg}},\ }\bibfield  {title} {\bibinfo {title} {\emph {Crossovers in
  Unitary Fermi Systems}},\ }in\ \href {\doibase 10.1007/978-3-642-21978-8_3}
  {\emph {\bibinfo {booktitle} {The BCS-BEC Crossover and the Unitary Fermi
  Gas}}},\ \bibinfo {editor} {edited by\ \bibinfo {editor} {\bibfnamefont
  {W.}~\bibnamefont {Zwerger}}}\ (\bibinfo  {publisher} {Springer},\ \bibinfo
  {address} {Berlin, Heidelberg},\ \bibinfo {year} {2012})\ pp.\ \bibinfo
  {pages} {49--97}\BibitemShut {NoStop}%
\bibitem [{\citenamefont {Petrov}(2003)}]{PhysRevA.67.010703}%
  \BibitemOpen
  \bibfield  {author} {\bibinfo {author} {\bibfnamefont {D.~S.}\ \bibnamefont
  {Petrov}},\ }\bibfield  {title} {\bibinfo {title} {\emph {Three-body problem
  in Fermi gases with short-range interparticle interaction}},\ }\href
  {\doibase 10.1103/PhysRevA.67.010703} {\bibfield  {journal} {\bibinfo
  {journal} {Phys. Rev. A}\ }\textbf {\bibinfo {volume} {67}},\ \bibinfo
  {pages} {010703} (\bibinfo {year} {2003})}\BibitemShut {NoStop}%
\bibitem [{\citenamefont {Petrov}\ \emph {et~al.}(2004)\citenamefont {Petrov},
  \citenamefont {Salomon},\ and\ \citenamefont
  {Shlyapnikov}}]{PhysRevLett.93.090404}%
  \BibitemOpen
  \bibfield  {author} {\bibinfo {author} {\bibfnamefont {D.~S.}\ \bibnamefont
  {Petrov}}, \bibinfo {author} {\bibfnamefont {C.}~\bibnamefont {Salomon}}, \
  and\ \bibinfo {author} {\bibfnamefont {G.~V.}\ \bibnamefont {Shlyapnikov}},\
  }\bibfield  {title} {\bibinfo {title} {\emph {Weakly Bound Dimers of
  Fermionic Atoms}},\ }\href {\doibase 10.1103/PhysRevLett.93.090404}
  {\bibfield  {journal} {\bibinfo  {journal} {Phys. Rev. Lett.}\ }\textbf
  {\bibinfo {volume} {93}},\ \bibinfo {pages} {090404} (\bibinfo {year}
  {2004})}\BibitemShut {NoStop}%
\bibitem [{\citenamefont {Ho}(2004)}]{Ho2004}%
  \BibitemOpen
  \bibfield  {author} {\bibinfo {author} {\bibfnamefont {T.-L.}\ \bibnamefont
  {Ho}},\ }\bibfield  {title} {\bibinfo {title} {\emph {Universal
  Thermodynamics of Degenerate Quantum Gases in the Unitarity Limit}},\ }\href
  {\doibase 10.1103/PhysRevLett.92.090402} {\bibfield  {journal} {\bibinfo
  {journal} {Phys. Rev. Lett.}\ }\textbf {\bibinfo {volume} {92}},\ \bibinfo
  {pages} {090402} (\bibinfo {year} {2004})}\BibitemShut {NoStop}%
\bibitem [{\citenamefont {Navon}\ \emph {et~al.}(2010)\citenamefont {Navon},
  \citenamefont {Nascimb{\`e}ne}, \citenamefont {Chevy},\ and\ \citenamefont
  {Salomon}}]{Navon2010}%
  \BibitemOpen
  \bibfield  {author} {\bibinfo {author} {\bibfnamefont {N.}~\bibnamefont
  {Navon}}, \bibinfo {author} {\bibfnamefont {S.}~\bibnamefont
  {Nascimb{\`e}ne}}, \bibinfo {author} {\bibfnamefont {F.}~\bibnamefont
  {Chevy}}, \ and\ \bibinfo {author} {\bibfnamefont {C.}~\bibnamefont
  {Salomon}},\ }\bibfield  {title} {\bibinfo {title} {\emph {The equation of
  state of a low-temperature Fermi gas with tunable interactions}},\ }\href
  {\doibase 10.1126/science.1187582} {\bibfield  {journal} {\bibinfo  {journal}
  {Science}\ }\textbf {\bibinfo {volume} {328}},\ \bibinfo {pages} {729}
  (\bibinfo {year} {2010})}\BibitemShut {NoStop}%
\bibitem [{\citenamefont {Horikoshi}\ \emph {et~al.}(2010)\citenamefont
  {Horikoshi}, \citenamefont {Nakajima}, \citenamefont {Ueda},\ and\
  \citenamefont {Mukaiyama}}]{Horikoshi2010}%
  \BibitemOpen
  \bibfield  {author} {\bibinfo {author} {\bibfnamefont {M.}~\bibnamefont
  {Horikoshi}}, \bibinfo {author} {\bibfnamefont {S.}~\bibnamefont {Nakajima}},
  \bibinfo {author} {\bibfnamefont {M.}~\bibnamefont {Ueda}}, \ and\ \bibinfo
  {author} {\bibfnamefont {T.}~\bibnamefont {Mukaiyama}},\ }\bibfield  {title}
  {\bibinfo {title} {\emph {Measurement of Universal Thermodynamic Functions
  for a Unitary Fermi Gas}},\ }\href {\doibase 10.1126/science.1183012}
  {\bibfield  {journal} {\bibinfo  {journal} {Science}\ }\textbf {\bibinfo
  {volume} {327}},\ \bibinfo {pages} {442} (\bibinfo {year}
  {2010})}\BibitemShut {NoStop}%
\bibitem [{\citenamefont {Ku}\ \emph {et~al.}(2012)\citenamefont {Ku},
  \citenamefont {Sommer}, \citenamefont {Cheuk},\ and\ \citenamefont
  {Zwierlein}}]{Ku2012}%
  \BibitemOpen
  \bibfield  {author} {\bibinfo {author} {\bibfnamefont {M.~J.~H.}\
  \bibnamefont {Ku}}, \bibinfo {author} {\bibfnamefont {A.~T.}\ \bibnamefont
  {Sommer}}, \bibinfo {author} {\bibfnamefont {L.~W.}\ \bibnamefont {Cheuk}}, \
  and\ \bibinfo {author} {\bibfnamefont {M.~W.}\ \bibnamefont {Zwierlein}},\
  }\bibfield  {title} {\bibinfo {title} {\emph {Revealing the Superfluid Lambda
  Transition in the Universal Thermodynamics of a Unitary Fermi Gas}},\ }\href
  {\doibase 10.1126/science.1214987} {\bibfield  {journal} {\bibinfo  {journal}
  {Science}\ }\textbf {\bibinfo {volume} {335}},\ \bibinfo {pages} {563}
  (\bibinfo {year} {2012})}\BibitemShut {NoStop}%
\bibitem [{\citenamefont {Van~Houcke}\ \emph {et~al.}(2012)\citenamefont
  {Van~Houcke}, \citenamefont {Werner}, \citenamefont {Kozik}, \citenamefont
  {Prokof'ev}, \citenamefont {Svistunov}, \citenamefont {Ku}, \citenamefont
  {Sommer}, \citenamefont {Cheuk}, \citenamefont {Schirotzek},\ and\
  \citenamefont {Zwierlein}}]{van2012feynman}%
  \BibitemOpen
  \bibfield  {author} {\bibinfo {author} {\bibfnamefont {K.}~\bibnamefont
  {Van~Houcke}}, \bibinfo {author} {\bibfnamefont {F.}~\bibnamefont {Werner}},
  \bibinfo {author} {\bibfnamefont {E.}~\bibnamefont {Kozik}}, \bibinfo
  {author} {\bibfnamefont {N.}~\bibnamefont {Prokof'ev}}, \bibinfo {author}
  {\bibfnamefont {B.}~\bibnamefont {Svistunov}}, \bibinfo {author}
  {\bibfnamefont {M.}~\bibnamefont {Ku}}, \bibinfo {author} {\bibfnamefont
  {A.}~\bibnamefont {Sommer}}, \bibinfo {author} {\bibfnamefont
  {L.}~\bibnamefont {Cheuk}}, \bibinfo {author} {\bibfnamefont
  {A.}~\bibnamefont {Schirotzek}}, \ and\ \bibinfo {author} {\bibfnamefont
  {M.}~\bibnamefont {Zwierlein}},\ }\bibfield  {title} {\bibinfo {title} {\emph
  {Feynman diagrams versus Fermi-gas Feynman emulator}},\ }\href {\doibase
  10.1038/NPHYS2273} {\bibfield  {journal} {\bibinfo  {journal} {Nature
  Physics}\ }\textbf {\bibinfo {volume} {8}},\ \bibinfo {pages} {366} (\bibinfo
  {year} {2012})}\BibitemShut {NoStop}%
\bibitem [{\citenamefont {Son}(2007)}]{son2007vis}%
  \BibitemOpen
  \bibfield  {author} {\bibinfo {author} {\bibfnamefont {D.~T.}\ \bibnamefont
  {Son}},\ }\bibfield  {title} {\bibinfo {title} {\emph {Vanishing Bulk
  Viscosities and Conformal Invariance of the Unitary Fermi Gas}},\ }\href
  {\doibase 10.1103/PhysRevLett.98.020604} {\bibfield  {journal} {\bibinfo
  {journal} {Phys. Rev. Lett.}\ }\textbf {\bibinfo {volume} {98}},\ \bibinfo
  {pages} {020604} (\bibinfo {year} {2007})}\BibitemShut {NoStop}%
\bibitem [{\citenamefont {Castin}\ and\ \citenamefont
  {Werner}(2012)}]{Castin2012}%
  \BibitemOpen
  \bibfield  {author} {\bibinfo {author} {\bibfnamefont {Y.}~\bibnamefont
  {Castin}}\ and\ \bibinfo {author} {\bibfnamefont {F.}~\bibnamefont
  {Werner}},\ }\bibfield  {title} {\bibinfo {title} {\emph {The Unitary Gas and
  its Symmetry Properties}},\ }in\ \href {\doibase 10.1007/978-3-642-21978-8_5}
  {\emph {\bibinfo {booktitle} {The BCS-BEC Crossover and the Unitary Fermi
  Gas}}},\ \bibinfo {editor} {edited by\ \bibinfo {editor} {\bibfnamefont
  {W.}~\bibnamefont {Zwerger}}}\ (\bibinfo  {publisher} {Springer},\ \bibinfo
  {address} {Berlin, Heidelberg},\ \bibinfo {year} {2012})\ pp.\ \bibinfo
  {pages} {127--191}\BibitemShut {NoStop}%
\bibitem [{\citenamefont {Cao}\ \emph {et~al.}(2011)\citenamefont {Cao},
  \citenamefont {Elliott}, \citenamefont {Joseph}, \citenamefont {Wu},
  \citenamefont {Petricka}, \citenamefont {Sch{\"a}fer},\ and\ \citenamefont
  {Thomas}}]{Cao2011}%
  \BibitemOpen
  \bibfield  {author} {\bibinfo {author} {\bibfnamefont {C.}~\bibnamefont
  {Cao}}, \bibinfo {author} {\bibfnamefont {E.}~\bibnamefont {Elliott}},
  \bibinfo {author} {\bibfnamefont {J.}~\bibnamefont {Joseph}}, \bibinfo
  {author} {\bibfnamefont {H.}~\bibnamefont {Wu}}, \bibinfo {author}
  {\bibfnamefont {J.}~\bibnamefont {Petricka}}, \bibinfo {author}
  {\bibfnamefont {T.}~\bibnamefont {Sch{\"a}fer}}, \ and\ \bibinfo {author}
  {\bibfnamefont {J.~E.}\ \bibnamefont {Thomas}},\ }\bibfield  {title}
  {\bibinfo {title} {\emph {Universal Quantum Viscosity in a Unitary Fermi
  Gas}},\ }\href {\doibase 10.1126/science.1195219} {\bibfield  {journal}
  {\bibinfo  {journal} {Science}\ }\textbf {\bibinfo {volume} {331}},\ \bibinfo
  {pages} {58} (\bibinfo {year} {2011})}\BibitemShut {NoStop}%
\bibitem [{\citenamefont {Hammond}\ \emph {et~al.}(1994)\citenamefont
  {Hammond}, \citenamefont {Lester~Jr.},\ and\ \citenamefont
  {Reynolds}}]{Hammond1994book}%
  \BibitemOpen
  \bibfield  {author} {\bibinfo {author} {\bibfnamefont {B.}~\bibnamefont
  {Hammond}}, \bibinfo {author} {\bibfnamefont {W.}~\bibnamefont {Lester~Jr.}},
  \ and\ \bibinfo {author} {\bibfnamefont {P.}~\bibnamefont {Reynolds}},\
  }\href {http://www.worldscientific.com/worldscibooks/10.1142/1170} {\emph
  {\bibinfo {title} {Monte Carlo Methods in Ab Initio Quantum Chemistry}}},\
  World Scientific Lecture and Course Notes in Chemistry\ (\bibinfo
  {publisher} {World Scientific},\ \bibinfo {year} {1994})\BibitemShut
  {NoStop}%
\bibitem [{\citenamefont {Randeria}\ \emph {et~al.}(2012)\citenamefont
  {Randeria}, \citenamefont {Zwerger},\ and\ \citenamefont
  {Zwierlein}}]{Randeria2012}%
  \BibitemOpen
  \bibfield  {author} {\bibinfo {author} {\bibfnamefont {M.}~\bibnamefont
  {Randeria}}, \bibinfo {author} {\bibfnamefont {W.}~\bibnamefont {Zwerger}}, \
  and\ \bibinfo {author} {\bibfnamefont {M.}~\bibnamefont {Zwierlein}},\
  }\bibinfo {title} {\emph {The BCS--BEC Crossover and the Unitary Fermi
  Gas}},\ in\ \href {\doibase 10.1007/978-3-642-21978-8_1} {\emph {\bibinfo
  {booktitle} {The BCS-BEC Crossover and the Unitary Fermi Gas}}},\ \bibinfo
  {editor} {edited by\ \bibinfo {editor} {\bibfnamefont {W.}~\bibnamefont
  {Zwerger}}}\ (\bibinfo  {publisher} {Springer Berlin Heidelberg},\ \bibinfo
  {address} {Berlin, Heidelberg},\ \bibinfo {year} {2012})\ pp.\ \bibinfo
  {pages} {1--32}\BibitemShut {NoStop}%
\bibitem [{\citenamefont {Strinati}(2012)}]{Strinati2012}%
  \BibitemOpen
  \bibfield  {author} {\bibinfo {author} {\bibfnamefont {G.~C.}\ \bibnamefont
  {Strinati}},\ }\bibinfo {title} {\emph {Pairing Fluctuations Approach to the
  BCS--BEC Crossover}},\ in\ \href {\doibase 10.1007/978-3-642-21978-8_4}
  {\emph {\bibinfo {booktitle} {The BCS-BEC Crossover and the Unitary Fermi
  Gas}}},\ \bibinfo {editor} {edited by\ \bibinfo {editor} {\bibfnamefont
  {W.}~\bibnamefont {Zwerger}}}\ (\bibinfo  {publisher} {Springer Berlin
  Heidelberg},\ \bibinfo {address} {Berlin, Heidelberg},\ \bibinfo {year}
  {2012})\ pp.\ \bibinfo {pages} {99--126}\BibitemShut {NoStop}%
\bibitem [{\citenamefont {Bulgac}\ \emph {et~al.}(2012)\citenamefont {Bulgac},
  \citenamefont {Forbes},\ and\ \citenamefont {Magierski}}]{Bulgac2012}%
  \BibitemOpen
  \bibfield  {author} {\bibinfo {author} {\bibfnamefont {A.}~\bibnamefont
  {Bulgac}}, \bibinfo {author} {\bibfnamefont {M.~M.}\ \bibnamefont {Forbes}},
  \ and\ \bibinfo {author} {\bibfnamefont {P.}~\bibnamefont {Magierski}},\
  }\bibinfo {title} {\emph {The Unitary Fermi Gas: From Monte Carlo to Density
  Functionals}},\ in\ \href {\doibase 10.1007/978-3-642-21978-8_9} {\emph
  {\bibinfo {booktitle} {The BCS-BEC Crossover and the Unitary Fermi Gas}}},\
  \bibinfo {editor} {edited by\ \bibinfo {editor} {\bibfnamefont
  {W.}~\bibnamefont {Zwerger}}}\ (\bibinfo  {publisher} {Springer Berlin
  Heidelberg},\ \bibinfo {address} {Berlin, Heidelberg},\ \bibinfo {year}
  {2012})\ pp.\ \bibinfo {pages} {305--373}\BibitemShut {NoStop}%
\bibitem [{\citenamefont {Chevy}\ and\ \citenamefont {Mora}(2010)}]{Chevy2010}%
  \BibitemOpen
  \bibfield  {author} {\bibinfo {author} {\bibfnamefont {F.}~\bibnamefont
  {Chevy}}\ and\ \bibinfo {author} {\bibfnamefont {C.}~\bibnamefont {Mora}},\
  }\bibfield  {title} {\bibinfo {title} {\emph {{Ultra-cold polarized Fermi
  gases}}},\ }\href {\doibase 10.1088/0034-4885/73/11/112401} {\bibfield
  {journal} {\bibinfo  {journal} {Rep. Progr. Phys.}\ }\textbf {\bibinfo
  {volume} {73}},\ \bibinfo {pages} {112401} (\bibinfo {year}
  {2010})}\BibitemShut {NoStop}%
\bibitem [{\citenamefont {Massignan}\ \emph {et~al.}(2014)\citenamefont
  {Massignan}, \citenamefont {Zaccanti},\ and\ \citenamefont
  {Bruun}}]{Massignan2014review}%
  \BibitemOpen
  \bibfield  {author} {\bibinfo {author} {\bibfnamefont {P.}~\bibnamefont
  {Massignan}}, \bibinfo {author} {\bibfnamefont {M.}~\bibnamefont {Zaccanti}},
  \ and\ \bibinfo {author} {\bibfnamefont {G.~M.}\ \bibnamefont {Bruun}},\
  }\bibfield  {title} {\bibinfo {title} {\emph {Polarons, dressed molecules and
  itinerant ferromagnetism in ultracold Fermi gases}},\ }\href
  {http://stacks.iop.org/0034-4885/77/i=3/a=034401} {\bibfield  {journal}
  {\bibinfo  {journal} {Rep. Progr. Phys.}\ }\textbf {\bibinfo {volume} {77}},\
  \bibinfo {pages} {034401} (\bibinfo {year} {2014})}\BibitemShut {NoStop}%
\bibitem [{\citenamefont {Liu}\ \emph {et~al.}(2009)\citenamefont {Liu},
  \citenamefont {Hu},\ and\ \citenamefont {Drummond}}]{Liu2009}%
  \BibitemOpen
  \bibfield  {author} {\bibinfo {author} {\bibfnamefont {X.-J.}\ \bibnamefont
  {Liu}}, \bibinfo {author} {\bibfnamefont {H.}~\bibnamefont {Hu}}, \ and\
  \bibinfo {author} {\bibfnamefont {P.~D.}\ \bibnamefont {Drummond}},\
  }\bibfield  {title} {\bibinfo {title} {\emph {Virial Expansion for a Strongly
  Correlated Fermi Gas}},\ }\href {\doibase 10.1103/PhysRevLett.102.160401}
  {\bibfield  {journal} {\bibinfo  {journal} {Phys. Rev. Lett.}\ }\textbf
  {\bibinfo {volume} {102}},\ \bibinfo {pages} {160401} (\bibinfo {year}
  {2009})}\BibitemShut {NoStop}%
\bibitem [{\citenamefont {Rakshit}\ \emph {et~al.}(2012)\citenamefont
  {Rakshit}, \citenamefont {Daily},\ and\ \citenamefont {Blume}}]{Rakshit2012}%
  \BibitemOpen
  \bibfield  {author} {\bibinfo {author} {\bibfnamefont {D.}~\bibnamefont
  {Rakshit}}, \bibinfo {author} {\bibfnamefont {K.~M.}\ \bibnamefont {Daily}},
  \ and\ \bibinfo {author} {\bibfnamefont {D.}~\bibnamefont {Blume}},\
  }\bibfield  {title} {\bibinfo {title} {\emph {Natural and unnatural parity
  states of small trapped equal-mass two-component Fermi gases at unitarity and
  fourth-order virial coefficient}},\ }\href {\doibase
  10.1103/PhysRevA.85.033634} {\bibfield  {journal} {\bibinfo  {journal} {Phys.
  Rev. A}\ }\textbf {\bibinfo {volume} {85}},\ \bibinfo {pages} {033634}
  (\bibinfo {year} {2012})}\BibitemShut {NoStop}%
\bibitem [{\citenamefont {Yan}\ and\ \citenamefont {Blume}(2016)}]{Yan2016}%
  \BibitemOpen
  \bibfield  {author} {\bibinfo {author} {\bibfnamefont {Y.}~\bibnamefont
  {Yan}}\ and\ \bibinfo {author} {\bibfnamefont {D.}~\bibnamefont {Blume}},\
  }\bibfield  {title} {\bibinfo {title} {\emph {Path-Integral Monte Carlo
  Determination of the Fourth-Order Virial Coefficient for a Unitary
  Two-Component Fermi Gas with Zero-Range Interactions}},\ }\href {\doibase
  10.1103/PhysRevLett.116.230401} {\bibfield  {journal} {\bibinfo  {journal}
  {Phys. Rev. Lett.}\ }\textbf {\bibinfo {volume} {116}},\ \bibinfo {pages}
  {230401} (\bibinfo {year} {2016})}\BibitemShut {NoStop}%
\bibitem [{\citenamefont {Endo}\ and\ \citenamefont {Castin}(2016)}]{Endo2016}%
  \BibitemOpen
  \bibfield  {author} {\bibinfo {author} {\bibfnamefont {S.}~\bibnamefont
  {Endo}}\ and\ \bibinfo {author} {\bibfnamefont {Y.}~\bibnamefont {Castin}},\
  }\bibfield  {title} {\bibinfo {title} {\emph {The interaction-sensitive
  states of a trapped two-component ideal Fermi gas and application to the
  virial expansion of the unitary Fermi gas}},\ }\href
  {http://stacks.iop.org/1751-8121/49/i=26/a=265301} {\bibfield  {journal}
  {\bibinfo  {journal} {Journal of Physics A: Mathematical and Theoretical}\
  }\textbf {\bibinfo {volume} {49}},\ \bibinfo {pages} {265301} (\bibinfo
  {year} {2016})}\BibitemShut {NoStop}%
\bibitem [{\citenamefont {Nascimb\`{e}ne}\ \emph {et~al.}(2010)\citenamefont
  {Nascimb\`{e}ne}, \citenamefont {Navon}, \citenamefont {Jiang}, \citenamefont
  {Chevy},\ and\ \citenamefont {Salomon}}]{Nascimbene2010}%
  \BibitemOpen
  \bibfield  {author} {\bibinfo {author} {\bibfnamefont {S.}~\bibnamefont
  {Nascimb\`{e}ne}}, \bibinfo {author} {\bibfnamefont {N.}~\bibnamefont
  {Navon}}, \bibinfo {author} {\bibfnamefont {K.~J.}\ \bibnamefont {Jiang}},
  \bibinfo {author} {\bibfnamefont {F.}~\bibnamefont {Chevy}}, \ and\ \bibinfo
  {author} {\bibfnamefont {C.}~\bibnamefont {Salomon}},\ }\bibfield  {title}
  {\bibinfo {title} {\emph {{Exploring the thermodynamics of a universal Fermi
  gas.}}},\ }\href {\doibase 10.1038/nature08814} {\bibfield  {journal}
  {\bibinfo  {journal} {Nature}\ }\textbf {\bibinfo {volume} {463}},\ \bibinfo
  {pages} {1057} (\bibinfo {year} {2010})}\BibitemShut {NoStop}%
\bibitem [{\citenamefont {Liu}(2013)}]{Liu2013}%
  \BibitemOpen
  \bibfield  {author} {\bibinfo {author} {\bibfnamefont {X.-J.}\ \bibnamefont
  {Liu}},\ }\bibfield  {title} {\bibinfo {title} {\emph {Virial expansion for a
  strongly correlated Fermi system and its application to ultracold atomic
  Fermi gases}},\ }\href {\doibase
  http://dx.doi.org/10.1016/j.physrep.2012.10.004} {\bibfield  {journal}
  {\bibinfo  {journal} {Physics Reports}\ }\textbf {\bibinfo {volume} {524}},\
  \bibinfo {pages} {37 } (\bibinfo {year} {2013})}\BibitemShut {NoStop}%
\bibitem [{\citenamefont {Wenz}\ \emph {et~al.}(2013)\citenamefont {Wenz},
  \citenamefont {Z{\"u}rn}, \citenamefont {Murmann}, \citenamefont {Brouzos},
  \citenamefont {Lompe},\ and\ \citenamefont {Jochim}}]{Wenz2013}%
  \BibitemOpen
  \bibfield  {author} {\bibinfo {author} {\bibfnamefont {A.}~\bibnamefont
  {Wenz}}, \bibinfo {author} {\bibfnamefont {G.}~\bibnamefont {Z{\"u}rn}},
  \bibinfo {author} {\bibfnamefont {S.}~\bibnamefont {Murmann}}, \bibinfo
  {author} {\bibfnamefont {I.}~\bibnamefont {Brouzos}}, \bibinfo {author}
  {\bibfnamefont {T.}~\bibnamefont {Lompe}}, \ and\ \bibinfo {author}
  {\bibfnamefont {S.}~\bibnamefont {Jochim}},\ }\bibfield  {title} {\bibinfo
  {title} {\emph {From few to many: observing the formation of a Fermi sea one
  atom at a time}},\ }\href {\doibase 10.1126/science.1240516} {\bibfield
  {journal} {\bibinfo  {journal} {Science}\ }\textbf {\bibinfo {volume}
  {342}},\ \bibinfo {pages} {457} (\bibinfo {year} {2013})}\BibitemShut
  {NoStop}%
\bibitem [{\citenamefont {Astrakharchik}\ and\ \citenamefont
  {Brouzos}(2013)}]{Astrakharchik2013}%
  \BibitemOpen
  \bibfield  {author} {\bibinfo {author} {\bibfnamefont {G.~E.}\ \bibnamefont
  {Astrakharchik}}\ and\ \bibinfo {author} {\bibfnamefont {I.}~\bibnamefont
  {Brouzos}},\ }\bibfield  {title} {\bibinfo {title} {\emph {Trapped
  one-dimensional ideal Fermi gas with a single impurity}},\ }\href {\doibase
  10.1103/PhysRevA.88.021602} {\bibfield  {journal} {\bibinfo  {journal} {Phys.
  Rev. A}\ }\textbf {\bibinfo {volume} {88}},\ \bibinfo {pages} {021602}
  (\bibinfo {year} {2013})}\BibitemShut {NoStop}%
\bibitem [{\citenamefont {Grining}\ \emph {et~al.}(2015)\citenamefont
  {Grining}, \citenamefont {Tomza}, \citenamefont {Lesiuk}, \citenamefont
  {Przybytek}, \citenamefont {Musia\l{}}, \citenamefont {Moszynski},
  \citenamefont {Lewenstein},\ and\ \citenamefont {Massignan}}]{Grining2015}%
  \BibitemOpen
  \bibfield  {author} {\bibinfo {author} {\bibfnamefont {T.}~\bibnamefont
  {Grining}}, \bibinfo {author} {\bibfnamefont {M.}~\bibnamefont {Tomza}},
  \bibinfo {author} {\bibfnamefont {M.}~\bibnamefont {Lesiuk}}, \bibinfo
  {author} {\bibfnamefont {M.}~\bibnamefont {Przybytek}}, \bibinfo {author}
  {\bibfnamefont {M.}~\bibnamefont {Musia\l{}}}, \bibinfo {author}
  {\bibfnamefont {R.}~\bibnamefont {Moszynski}}, \bibinfo {author}
  {\bibfnamefont {M.}~\bibnamefont {Lewenstein}}, \ and\ \bibinfo {author}
  {\bibfnamefont {P.}~\bibnamefont {Massignan}},\ }\bibfield  {title} {\bibinfo
  {title} {\emph {Crossover between few and many fermions in a harmonic
  trap}},\ }\href {\doibase 10.1103/PhysRevA.92.061601} {\bibfield  {journal}
  {\bibinfo  {journal} {Phys. Rev. A}\ }\textbf {\bibinfo {volume} {92}},\
  \bibinfo {pages} {061601} (\bibinfo {year} {2015})}\BibitemShut {NoStop}%
\bibitem [{\citenamefont {Blume}(2012)}]{Blume2012}%
  \BibitemOpen
  \bibfield  {author} {\bibinfo {author} {\bibfnamefont {D.}~\bibnamefont
  {Blume}},\ }\bibfield  {title} {\bibinfo {title} {\emph {Few-body physics
  with ultracold atomic and molecular systems in traps}},\ }\href
  {http://stacks.iop.org/0034-4885/75/i=4/a=046401} {\bibfield  {journal}
  {\bibinfo  {journal} {Reports on Progress in Physics}\ }\textbf {\bibinfo
  {volume} {75}},\ \bibinfo {pages} {046401} (\bibinfo {year}
  {2012})}\BibitemShut {NoStop}%
\bibitem [{\citenamefont {{Valtolina}}\ \emph {et~al.}(2016)\citenamefont
  {{Valtolina}}, \citenamefont {{Scazza}}, \citenamefont {{Amico}},
  \citenamefont {{Burchianti}}, \citenamefont {{Recati}}, \citenamefont
  {{Enss}}, \citenamefont {{Inguscio}}, \citenamefont {{Zaccanti}},\ and\
  \citenamefont {{Roati}}}]{Valtolina2016}%
  \BibitemOpen
  \bibfield  {author} {\bibinfo {author} {\bibfnamefont {G.}~\bibnamefont
  {{Valtolina}}}, \bibinfo {author} {\bibfnamefont {F.}~\bibnamefont
  {{Scazza}}}, \bibinfo {author} {\bibfnamefont {A.}~\bibnamefont {{Amico}}},
  \bibinfo {author} {\bibfnamefont {A.}~\bibnamefont {{Burchianti}}}, \bibinfo
  {author} {\bibfnamefont {A.}~\bibnamefont {{Recati}}}, \bibinfo {author}
  {\bibfnamefont {T.}~\bibnamefont {{Enss}}}, \bibinfo {author} {\bibfnamefont
  {M.}~\bibnamefont {{Inguscio}}}, \bibinfo {author} {\bibfnamefont
  {M.}~\bibnamefont {{Zaccanti}}}, \ and\ \bibinfo {author} {\bibfnamefont
  {G.}~\bibnamefont {{Roati}}},\ }\bibfield  {title} {\bibinfo {title} {\emph
  {{Evidence for ferromagnetic instability in a repulsive Fermi gas of
  ultracold atoms}}},\ }\href
  {http://adsabs.harvard.edu/abs/2016arXiv160507850V} {\bibfield  {journal}
  {\bibinfo  {journal} {arXiv:1605.07850}\ } (\bibinfo {year}
  {2016})}\BibitemShut {NoStop}%
\bibitem [{\citenamefont {Chevy}\ and\ \citenamefont
  {Salomon}(2012)}]{Chevy2012}%
  \BibitemOpen
  \bibfield  {author} {\bibinfo {author} {\bibfnamefont {F.}~\bibnamefont
  {Chevy}}\ and\ \bibinfo {author} {\bibfnamefont {C.}~\bibnamefont
  {Salomon}},\ }\bibinfo {title} {\emph {Thermodynamics of Fermi Gases}},\ in\
  \href {\doibase 10.1007/978-3-642-21978-8_11} {\emph {\bibinfo {booktitle}
  {The BCS-BEC Crossover and the Unitary Fermi Gas}}},\ \bibinfo {editor}
  {edited by\ \bibinfo {editor} {\bibfnamefont {W.}~\bibnamefont {Zwerger}}}\
  (\bibinfo  {publisher} {Springer Berlin Heidelberg},\ \bibinfo {address}
  {Berlin, Heidelberg},\ \bibinfo {year} {2012})\ pp.\ \bibinfo {pages}
  {407--446}\BibitemShut {NoStop}%
\bibitem [{\citenamefont {Schirotzek}\ \emph {et~al.}(2008)\citenamefont
  {Schirotzek}, \citenamefont {Shin}, \citenamefont {Schunck},\ and\
  \citenamefont {Ketterle}}]{Schirotzek2008}%
  \BibitemOpen
  \bibfield  {author} {\bibinfo {author} {\bibfnamefont {A.}~\bibnamefont
  {Schirotzek}}, \bibinfo {author} {\bibfnamefont {Y.-i.}\ \bibnamefont
  {Shin}}, \bibinfo {author} {\bibfnamefont {C.~H.}\ \bibnamefont {Schunck}}, \
  and\ \bibinfo {author} {\bibfnamefont {W.}~\bibnamefont {Ketterle}},\
  }\bibfield  {title} {\bibinfo {title} {\emph {Determination of the Superfluid
  Gap in Atomic Fermi Gases by Quasiparticle Spectroscopy}},\ }\href {\doibase
  10.1103/PhysRevLett.101.140403} {\bibfield  {journal} {\bibinfo  {journal}
  {Phys. Rev. Lett.}\ }\textbf {\bibinfo {volume} {101}},\ \bibinfo {pages}
  {140403} (\bibinfo {year} {2008})}\BibitemShut {NoStop}%
\bibitem [{\citenamefont {Bruun}\ and\ \citenamefont
  {Heiselberg}(2002)}]{Bruun2002}%
  \BibitemOpen
  \bibfield  {author} {\bibinfo {author} {\bibfnamefont {G.~M.}\ \bibnamefont
  {Bruun}}\ and\ \bibinfo {author} {\bibfnamefont {H.}~\bibnamefont
  {Heiselberg}},\ }\bibfield  {title} {\bibinfo {title} {\emph {Cooper pairing
  and single-particle properties of trapped Fermi gases}},\ }\href {\doibase
  10.1103/PhysRevA.65.053407} {\bibfield  {journal} {\bibinfo  {journal} {Phys.
  Rev. A}\ }\textbf {\bibinfo {volume} {65}},\ \bibinfo {pages} {053407}
  (\bibinfo {year} {2002})}\BibitemShut {NoStop}%
\bibitem [{\citenamefont {Leggett}(1980)}]{Leggett1980}%
  \BibitemOpen
  \bibfield  {author} {\bibinfo {author} {\bibfnamefont {A.}~\bibnamefont
  {Leggett}},\ }\bibfield  {title} {\bibinfo {title} {\emph {Diatomic molecules
  and cooper pairs}},\ }in\ \href {\doibase 10.1007/BFb0120125} {\emph
  {\bibinfo {booktitle} {Modern Trends in the Theory of Condensed Matter}}},\
  \bibinfo {series} {Lecture Notes in Physics}, Vol.\ \bibinfo {volume} {115},\
  \bibinfo {editor} {edited by\ \bibinfo {editor} {\bibfnamefont
  {A.}~\bibnamefont {P\c{e}kalski}}\ and\ \bibinfo {editor} {\bibfnamefont
  {J.}~\bibnamefont {Przystawa}}}\ (\bibinfo  {publisher} {Springer Berlin
  Heidelberg},\ \bibinfo {year} {1980})\ pp.\ \bibinfo {pages}
  {13--27}\BibitemShut {NoStop}%
\bibitem [{\citenamefont {Levinsen}\ and\ \citenamefont
  {Parish}(2015)}]{Levinsen2015review}%
  \BibitemOpen
  \bibfield  {author} {\bibinfo {author} {\bibfnamefont {J.}~\bibnamefont
  {Levinsen}}\ and\ \bibinfo {author} {\bibfnamefont {M.~M.}\ \bibnamefont
  {Parish}},\ }\bibinfo {title} {\emph {Strongly interacting two-dimensional
  Fermi gases}},\ in\ \href {\doibase 10.1142/9789814667746_0001} {\emph
  {\bibinfo {booktitle} {Annual Review of Cold Atoms and Molecules}}}\
  (\bibinfo  {publisher} {World Scientific},\ \bibinfo {year} {2015})\ pp.\
  \bibinfo {pages} {1--75}\BibitemShut {NoStop}%
\bibitem [{\citenamefont {Lobo}\ \emph {et~al.}(2006)\citenamefont {Lobo},
  \citenamefont {Recati}, \citenamefont {Giorgini},\ and\ \citenamefont
  {Stringari}}]{Lobo2006}%
  \BibitemOpen
  \bibfield  {author} {\bibinfo {author} {\bibfnamefont {C.}~\bibnamefont
  {Lobo}}, \bibinfo {author} {\bibfnamefont {A.}~\bibnamefont {Recati}},
  \bibinfo {author} {\bibfnamefont {S.}~\bibnamefont {Giorgini}}, \ and\
  \bibinfo {author} {\bibfnamefont {S.}~\bibnamefont {Stringari}},\ }\bibfield
  {title} {\bibinfo {title} {\emph {Normal State of a Polarized Fermi Gas at
  Unitarity}},\ }\href {\doibase 10.1103/PhysRevLett.97.200403} {\bibfield
  {journal} {\bibinfo  {journal} {Phys. Rev. Lett.}\ }\textbf {\bibinfo
  {volume} {97}},\ \bibinfo {pages} {200403} (\bibinfo {year}
  {2006})}\BibitemShut {NoStop}%
\bibitem [{\citenamefont {Pilati}\ and\ \citenamefont
  {Giorgini}(2008)}]{Pilati2008}%
  \BibitemOpen
  \bibfield  {author} {\bibinfo {author} {\bibfnamefont {S.}~\bibnamefont
  {Pilati}}\ and\ \bibinfo {author} {\bibfnamefont {S.}~\bibnamefont
  {Giorgini}},\ }\bibfield  {title} {\bibinfo {title} {\emph {Phase Separation
  in a Polarized Fermi Gas at Zero Temperature}},\ }\href {\doibase
  10.1103/PhysRevLett.100.030401} {\bibfield  {journal} {\bibinfo  {journal}
  {Phys. Rev. Lett.}\ }\textbf {\bibinfo {volume} {100}},\ \bibinfo {pages}
  {030401} (\bibinfo {year} {2008})}\BibitemShut {NoStop}%
\bibitem [{\citenamefont {Chevy}(2006)}]{Chevy2006}%
  \BibitemOpen
  \bibfield  {author} {\bibinfo {author} {\bibfnamefont {F.}~\bibnamefont
  {Chevy}},\ }\bibfield  {title} {\bibinfo {title} {\emph {Universal phase
  diagram of a strongly interacting Fermi gas with unbalanced spin
  populations}},\ }\href {\doibase 10.1103/PhysRevA.74.063628} {\bibfield
  {journal} {\bibinfo  {journal} {Phys. Rev. A}\ }\textbf {\bibinfo {volume}
  {74}},\ \bibinfo {pages} {063628} (\bibinfo {year} {2006})}\BibitemShut
  {NoStop}%
\bibitem [{\citenamefont {Combescot}\ \emph {et~al.}(2007)\citenamefont
  {Combescot}, \citenamefont {Recati}, \citenamefont {Lobo},\ and\
  \citenamefont {Chevy}}]{Combescot2007}%
  \BibitemOpen
  \bibfield  {author} {\bibinfo {author} {\bibfnamefont {R.}~\bibnamefont
  {Combescot}}, \bibinfo {author} {\bibfnamefont {A.}~\bibnamefont {Recati}},
  \bibinfo {author} {\bibfnamefont {C.}~\bibnamefont {Lobo}}, \ and\ \bibinfo
  {author} {\bibfnamefont {F.}~\bibnamefont {Chevy}},\ }\bibfield  {title}
  {\bibinfo {title} {\emph {Normal State of Highly Polarized Fermi Gases:
  Simple Many-Body Approaches}},\ }\href {\doibase
  10.1103/PhysRevLett.98.180402} {\bibfield  {journal} {\bibinfo  {journal}
  {Phys. Rev. Lett.}\ }\textbf {\bibinfo {volume} {98}},\ \bibinfo {pages}
  {180402} (\bibinfo {year} {2007})}\BibitemShut {NoStop}%
\bibitem [{\citenamefont {Prokof'ev}\ and\ \citenamefont
  {Svistunov}(2008)}]{Prokofev2008}%
  \BibitemOpen
  \bibfield  {author} {\bibinfo {author} {\bibfnamefont {N.}~\bibnamefont
  {Prokof'ev}}\ and\ \bibinfo {author} {\bibfnamefont {B.}~\bibnamefont
  {Svistunov}},\ }\bibfield  {title} {\bibinfo {title} {\emph {Fermi-polaron
  problem: Diagrammatic Monte Carlo method for divergent sign-alternating
  series}},\ }\href {\doibase 10.1103/PhysRevB.77.020408} {\bibfield  {journal}
  {\bibinfo  {journal} {Phys. Rev. B}\ }\textbf {\bibinfo {volume} {77}},\
  \bibinfo {pages} {020408} (\bibinfo {year} {2008})}\BibitemShut {NoStop}%
\bibitem [{\citenamefont {Combescot}\ and\ \citenamefont
  {Giraud}(2008)}]{Combescot2008}%
  \BibitemOpen
  \bibfield  {author} {\bibinfo {author} {\bibfnamefont {R.}~\bibnamefont
  {Combescot}}\ and\ \bibinfo {author} {\bibfnamefont {S.}~\bibnamefont
  {Giraud}},\ }\bibfield  {title} {\bibinfo {title} {\emph {Normal State of
  Highly Polarized Fermi Gases: Full Many-Body Treatment}},\ }\href {\doibase
  10.1103/PhysRevLett.101.050404} {\bibfield  {journal} {\bibinfo  {journal}
  {Phys. Rev. Lett.}\ }\textbf {\bibinfo {volume} {101}},\ \bibinfo {pages}
  {050404} (\bibinfo {year} {2008})}\BibitemShut {NoStop}%
\bibitem [{\citenamefont {Vlietinck}\ \emph {et~al.}(2013)\citenamefont
  {Vlietinck}, \citenamefont {Ryckebusch},\ and\ \citenamefont
  {Van~Houcke}}]{Vlietinck2013}%
  \BibitemOpen
  \bibfield  {author} {\bibinfo {author} {\bibfnamefont {J.}~\bibnamefont
  {Vlietinck}}, \bibinfo {author} {\bibfnamefont {J.}~\bibnamefont
  {Ryckebusch}}, \ and\ \bibinfo {author} {\bibfnamefont {K.}~\bibnamefont
  {Van~Houcke}},\ }\bibfield  {title} {\bibinfo {title} {\emph {Quasiparticle
  properties of an impurity in a Fermi gas}},\ }\href {\doibase
  10.1103/PhysRevB.87.115133} {\bibfield  {journal} {\bibinfo  {journal} {Phys.
  Rev. B}\ }\textbf {\bibinfo {volume} {87}},\ \bibinfo {pages} {115133}
  (\bibinfo {year} {2013})}\BibitemShut {NoStop}%
\bibitem [{\citenamefont {Pethick}\ and\ \citenamefont
  {Smith}(2008)}]{PethickSmith2008book}%
  \BibitemOpen
  \bibfield  {author} {\bibinfo {author} {\bibfnamefont {C.~J.}\ \bibnamefont
  {Pethick}}\ and\ \bibinfo {author} {\bibfnamefont {H.}~\bibnamefont
  {Smith}},\ }\href@noop {} {\emph {\bibinfo {title} {Bose-Einstein
  Condensation in Dilute Gases}}}\ (\bibinfo  {publisher} {Cambridge University
  Press},\ \bibinfo {address} {Cambridge},\ \bibinfo {year} {2008})\BibitemShut
  {NoStop}%
\bibitem [{\citenamefont {Werner}\ and\ \citenamefont
  {Castin}(2012)}]{Werner2012a}%
  \BibitemOpen
  \bibfield  {author} {\bibinfo {author} {\bibfnamefont {F.}~\bibnamefont
  {Werner}}\ and\ \bibinfo {author} {\bibfnamefont {Y.}~\bibnamefont
  {Castin}},\ }\bibfield  {title} {\bibinfo {title} {\emph {General relations
  for quantum gases in two and three dimensions: Two-component fermions}},\
  }\href {\doibase 10.1103/PhysRevA.86.013626} {\bibfield  {journal} {\bibinfo
  {journal} {Phys. Rev. A}\ }\textbf {\bibinfo {volume} {86}},\ \bibinfo
  {pages} {013626} (\bibinfo {year} {2012})}\BibitemShut {NoStop}%
\bibitem [{\citenamefont {Werner}\ \emph {et~al.}(2009)\citenamefont {Werner},
  \citenamefont {Tarruell},\ and\ \citenamefont {Castin}}]{Werner2009}%
  \BibitemOpen
  \bibfield  {author} {\bibinfo {author} {\bibfnamefont {F.}~\bibnamefont
  {Werner}}, \bibinfo {author} {\bibfnamefont {L.}~\bibnamefont {Tarruell}}, \
  and\ \bibinfo {author} {\bibfnamefont {Y.}~\bibnamefont {Castin}},\
  }\bibfield  {title} {\bibinfo {title} {\emph {{Number of closed-channel
  molecules in the BEC-BCS crossover}}},\ }\href {\doibase
  10.1140/epjb/e2009-00040-8} {\bibfield  {journal} {\bibinfo  {journal} {The
  European Physical Journal B}\ }\textbf {\bibinfo {volume} {68}},\ \bibinfo
  {pages} {401} (\bibinfo {year} {2009})}\BibitemShut {NoStop}%
\bibitem [{\citenamefont {Hoinka}\ \emph {et~al.}(2013)\citenamefont {Hoinka},
  \citenamefont {Lingham}, \citenamefont {Fenech}, \citenamefont {Hu},
  \citenamefont {Vale}, \citenamefont {Drut},\ and\ \citenamefont
  {Gandolfi}}]{Hoinka2013}%
  \BibitemOpen
  \bibfield  {author} {\bibinfo {author} {\bibfnamefont {S.}~\bibnamefont
  {Hoinka}}, \bibinfo {author} {\bibfnamefont {M.}~\bibnamefont {Lingham}},
  \bibinfo {author} {\bibfnamefont {K.}~\bibnamefont {Fenech}}, \bibinfo
  {author} {\bibfnamefont {H.}~\bibnamefont {Hu}}, \bibinfo {author}
  {\bibfnamefont {C.~J.}\ \bibnamefont {Vale}}, \bibinfo {author}
  {\bibfnamefont {J.~E.}\ \bibnamefont {Drut}}, \ and\ \bibinfo {author}
  {\bibfnamefont {S.}~\bibnamefont {Gandolfi}},\ }\bibfield  {title} {\bibinfo
  {title} {\emph {Precise Determination of the Structure Factor and Contact in
  a Unitary Fermi Gas}},\ }\href {\doibase 10.1103/PhysRevLett.110.055305}
  {\bibfield  {journal} {\bibinfo  {journal} {Phys. Rev. Lett.}\ }\textbf
  {\bibinfo {volume} {110}},\ \bibinfo {pages} {055305} (\bibinfo {year}
  {2013})}\BibitemShut {NoStop}%
\bibitem [{\citenamefont {Busch}\ \emph {et~al.}(1998)\citenamefont {Busch},
  \citenamefont {Englert}, \citenamefont {Rza{\.{z}}ewski},\ and\ \citenamefont
  {Wilkens}}]{Busch1998}%
  \BibitemOpen
  \bibfield  {author} {\bibinfo {author} {\bibfnamefont {T.}~\bibnamefont
  {Busch}}, \bibinfo {author} {\bibfnamefont {B.-G.}\ \bibnamefont {Englert}},
  \bibinfo {author} {\bibfnamefont {K.}~\bibnamefont {Rza{\.{z}}ewski}}, \ and\
  \bibinfo {author} {\bibfnamefont {M.}~\bibnamefont {Wilkens}},\ }\bibfield
  {title} {\bibinfo {title} {\emph {Two Cold Atoms in a Harmonic Trap}},\
  }\href {\doibase 10.1023/A:1018705520999} {\bibfield  {journal} {\bibinfo
  {journal} {Foundations of Physics}\ }\textbf {\bibinfo {volume} {28}},\
  \bibinfo {pages} {549} (\bibinfo {year} {1998})}\BibitemShut {NoStop}%
\bibitem [{\citenamefont {Yin}\ and\ \citenamefont {Blume}(2015)}]{Yin2015}%
  \BibitemOpen
  \bibfield  {author} {\bibinfo {author} {\bibfnamefont {X.~Y.}\ \bibnamefont
  {Yin}}\ and\ \bibinfo {author} {\bibfnamefont {D.}~\bibnamefont {Blume}},\
  }\bibfield  {title} {\bibinfo {title} {\emph {Trapped unitary two-component
  Fermi gases with up to ten particles}},\ }\href {\doibase
  10.1103/PhysRevA.92.013608} {\bibfield  {journal} {\bibinfo  {journal} {Phys.
  Rev. A}\ }\textbf {\bibinfo {volume} {92}},\ \bibinfo {pages} {013608}
  (\bibinfo {year} {2015})}\BibitemShut {NoStop}%
\bibitem [{\citenamefont {Werner}\ and\ \citenamefont
  {Castin}(2006)}]{Werner2006}%
  \BibitemOpen
  \bibfield  {author} {\bibinfo {author} {\bibfnamefont {F.}~\bibnamefont
  {Werner}}\ and\ \bibinfo {author} {\bibfnamefont {Y.}~\bibnamefont
  {Castin}},\ }\bibfield  {title} {\bibinfo {title} {\emph {Unitary Quantum
  Three-Body Problem in a Harmonic Trap}},\ }\href {\doibase
  10.1103/PhysRevLett.97.150401} {\bibfield  {journal} {\bibinfo  {journal}
  {Phys. Rev. Lett.}\ }\textbf {\bibinfo {volume} {97}},\ \bibinfo {pages}
  {150401} (\bibinfo {year} {2006})}\BibitemShut {NoStop}%
\bibitem [{\citenamefont {Daily}\ and\ \citenamefont
  {Blume}(2010)}]{Daily2010}%
  \BibitemOpen
  \bibfield  {author} {\bibinfo {author} {\bibfnamefont {K.~M.}\ \bibnamefont
  {Daily}}\ and\ \bibinfo {author} {\bibfnamefont {D.}~\bibnamefont {Blume}},\
  }\bibfield  {title} {\bibinfo {title} {\emph {Energy spectrum of harmonically
  trapped two-component Fermi gases: Three- and four-particle problem}},\
  }\href {\doibase 10.1103/PhysRevA.81.053615} {\bibfield  {journal} {\bibinfo
  {journal} {Phys. Rev. A}\ }\textbf {\bibinfo {volume} {81}},\ \bibinfo
  {pages} {053615} (\bibinfo {year} {2010})}\BibitemShut {NoStop}%
\bibitem [{\citenamefont {Blume}\ and\ \citenamefont
  {Daily}(2010)}]{BlumeDaily2010}%
  \BibitemOpen
  \bibfield  {author} {\bibinfo {author} {\bibfnamefont {D.}~\bibnamefont
  {Blume}}\ and\ \bibinfo {author} {\bibfnamefont {K.~M.}\ \bibnamefont
  {Daily}},\ }\bibfield  {title} {\bibinfo {title} {\emph {Few-body resonances
  of unequal-mass systems with infinite interspecies two-body $s$-wave
  scattering length}},\ }\href {\doibase 10.1103/PhysRevA.82.063612} {\bibfield
   {journal} {\bibinfo  {journal} {Phys. Rev. A}\ }\textbf {\bibinfo {volume}
  {82}},\ \bibinfo {pages} {063612} (\bibinfo {year} {2010})}\BibitemShut
  {NoStop}%
\bibitem [{\citenamefont {Blume}\ and\ \citenamefont
  {Daily}(2011)}]{Blume2011}%
  \BibitemOpen
  \bibfield  {author} {\bibinfo {author} {\bibfnamefont {D.}~\bibnamefont
  {Blume}}\ and\ \bibinfo {author} {\bibfnamefont {K.}~\bibnamefont {Daily}},\
  }\bibfield  {title} {\bibinfo {title} {\emph {Trapped two-component Fermi
  gases with up to six particles: Energetics, structural properties, and
  molecular condensate fraction}},\ }\href {\doibase
  http://dx.doi.org/10.1016/j.crhy.2010.11.010} {\bibfield  {journal} {\bibinfo
   {journal} {Comptes Rendus Physique}\ }\textbf {\bibinfo {volume} {12}},\
  \bibinfo {pages} {86 } (\bibinfo {year} {2011})}\BibitemShut {NoStop}%
\bibitem [{\citenamefont {Bradly}\ \emph {et~al.}(2014)\citenamefont {Bradly},
  \citenamefont {Mulkerin}, \citenamefont {Martin},\ and\ \citenamefont
  {Quiney}}]{Bradly2014}%
  \BibitemOpen
  \bibfield  {author} {\bibinfo {author} {\bibfnamefont {C.~J.}\ \bibnamefont
  {Bradly}}, \bibinfo {author} {\bibfnamefont {B.~C.}\ \bibnamefont
  {Mulkerin}}, \bibinfo {author} {\bibfnamefont {A.~M.}\ \bibnamefont
  {Martin}}, \ and\ \bibinfo {author} {\bibfnamefont {H.~M.}\ \bibnamefont
  {Quiney}},\ }\bibfield  {title} {\bibinfo {title} {\emph {Coupled-pair
  approach for strongly interacting trapped fermionic atoms}},\ }\href
  {\doibase 10.1103/PhysRevA.90.023626} {\bibfield  {journal} {\bibinfo
  {journal} {Phys. Rev. A}\ }\textbf {\bibinfo {volume} {90}},\ \bibinfo
  {pages} {023626} (\bibinfo {year} {2014})}\BibitemShut {NoStop}%
\bibitem [{\citenamefont {Blume}\ \emph {et~al.}(2007)\citenamefont {Blume},
  \citenamefont {von Stecher},\ and\ \citenamefont {Greene}}]{Blume2007}%
  \BibitemOpen
  \bibfield  {author} {\bibinfo {author} {\bibfnamefont {D.}~\bibnamefont
  {Blume}}, \bibinfo {author} {\bibfnamefont {J.}~\bibnamefont {von Stecher}},
  \ and\ \bibinfo {author} {\bibfnamefont {C.~H.}\ \bibnamefont {Greene}},\
  }\bibfield  {title} {\bibinfo {title} {\emph {Universal Properties of a
  Trapped Two-Component Fermi Gas at Unitarity}},\ }\href {\doibase
  10.1103/PhysRevLett.99.233201} {\bibfield  {journal} {\bibinfo  {journal}
  {Phys. Rev. Lett.}\ }\textbf {\bibinfo {volume} {99}},\ \bibinfo {pages}
  {233201} (\bibinfo {year} {2007})}\BibitemShut {NoStop}%
\bibitem [{\citenamefont {Bulgac}(2007)}]{Bulgac2007}%
  \BibitemOpen
  \bibfield  {author} {\bibinfo {author} {\bibfnamefont {A.}~\bibnamefont
  {Bulgac}},\ }\bibfield  {title} {\bibinfo {title} {\emph
  {Local-density-functional theory for superfluid fermionic systems: The
  unitary gas}},\ }\href {\doibase 10.1103/PhysRevA.76.040502} {\bibfield
  {journal} {\bibinfo  {journal} {Phys. Rev. A}\ }\textbf {\bibinfo {volume}
  {76}},\ \bibinfo {pages} {040502} (\bibinfo {year} {2007})}\BibitemShut
  {NoStop}%
\bibitem [{\citenamefont {J\'auregui}\ \emph {et~al.}(2007)\citenamefont
  {J\'auregui}, \citenamefont {Paredes},\ and\ \citenamefont
  {S\'anchez}}]{Jauregui2007}%
  \BibitemOpen
  \bibfield  {author} {\bibinfo {author} {\bibfnamefont {R.}~\bibnamefont
  {J\'auregui}}, \bibinfo {author} {\bibfnamefont {R.}~\bibnamefont {Paredes}},
  \ and\ \bibinfo {author} {\bibfnamefont {G.~T.}\ \bibnamefont {S\'anchez}},\
  }\bibfield  {title} {\bibinfo {title} {\emph {BEC-BCS crossover of a trapped
  Fermi gas without using the local density approximation}},\ }\href {\doibase
  10.1103/PhysRevA.76.011604} {\bibfield  {journal} {\bibinfo  {journal} {Phys.
  Rev. A}\ }\textbf {\bibinfo {volume} {76}},\ \bibinfo {pages} {011604}
  (\bibinfo {year} {2007})}\BibitemShut {NoStop}%
\bibitem [{\citenamefont {Chang}\ and\ \citenamefont
  {Bertsch}(2007)}]{Chang2007}%
  \BibitemOpen
  \bibfield  {author} {\bibinfo {author} {\bibfnamefont {S.~Y.}\ \bibnamefont
  {Chang}}\ and\ \bibinfo {author} {\bibfnamefont {G.~F.}\ \bibnamefont
  {Bertsch}},\ }\bibfield  {title} {\bibinfo {title} {\emph {Unitary Fermi gas
  in a harmonic trap}},\ }\href {\doibase 10.1103/PhysRevA.76.021603}
  {\bibfield  {journal} {\bibinfo  {journal} {Phys. Rev. A}\ }\textbf {\bibinfo
  {volume} {76}},\ \bibinfo {pages} {021603} (\bibinfo {year}
  {2007})}\BibitemShut {NoStop}%
\bibitem [{\citenamefont {von Stecher}\ \emph {et~al.}(2008)\citenamefont {von
  Stecher}, \citenamefont {Greene},\ and\ \citenamefont
  {Blume}}]{vonStecher2008}%
  \BibitemOpen
  \bibfield  {author} {\bibinfo {author} {\bibfnamefont {J.}~\bibnamefont {von
  Stecher}}, \bibinfo {author} {\bibfnamefont {C.~H.}\ \bibnamefont {Greene}},
  \ and\ \bibinfo {author} {\bibfnamefont {D.}~\bibnamefont {Blume}},\
  }\bibfield  {title} {\bibinfo {title} {\emph {Energetics and structural
  properties of trapped two-component Fermi gases}},\ }\href {\doibase
  10.1103/PhysRevA.77.043619} {\bibfield  {journal} {\bibinfo  {journal} {Phys.
  Rev. A}\ }\textbf {\bibinfo {volume} {77}},\ \bibinfo {pages} {043619}
  (\bibinfo {year} {2008})}\BibitemShut {NoStop}%
\bibitem [{\citenamefont {Zubarev}\ and\ \citenamefont
  {Zoubarev}(2009)}]{Zubarev2009}%
  \BibitemOpen
  \bibfield  {author} {\bibinfo {author} {\bibfnamefont {A.~L.}\ \bibnamefont
  {Zubarev}}\ and\ \bibinfo {author} {\bibfnamefont {M.}~\bibnamefont
  {Zoubarev}},\ }\bibfield  {title} {\bibinfo {title} {\emph {On the kinetic
  energy of unitary Fermi gas in a harmonic trap}},\ }\href
  {http://stacks.iop.org/0295-5075/87/i=3/a=33001} {\bibfield  {journal}
  {\bibinfo  {journal} {EPL (Europhysics Letters)}\ }\textbf {\bibinfo {volume}
  {87}},\ \bibinfo {pages} {33001} (\bibinfo {year} {2009})}\BibitemShut
  {NoStop}%
\bibitem [{\citenamefont {{Nicholson}}\ \emph {et~al.}(2010)\citenamefont
  {{Nicholson}}, \citenamefont {{Endres}}, \citenamefont {{Kaplan}},\ and\
  \citenamefont {{Lee}}}]{Nicholson2010}%
  \BibitemOpen
  \bibfield  {author} {\bibinfo {author} {\bibfnamefont {A.}~\bibnamefont
  {{Nicholson}}}, \bibinfo {author} {\bibfnamefont {M.}~\bibnamefont
  {{Endres}}}, \bibinfo {author} {\bibfnamefont {D.~B.}\ \bibnamefont
  {{Kaplan}}}, \ and\ \bibinfo {author} {\bibfnamefont {J.~W.}\ \bibnamefont
  {{Lee}}},\ }\bibfield  {title} {\bibinfo {title} {\emph {{Lattice Study of
  Trapped Fermions at Unitarity}}},\ }in\ \href@noop {} {\emph {\bibinfo
  {booktitle} {Proceedings of The XXVIII International Symposium on Lattice
  Field Theory. June 14-19,2010. Villasimius, Sardinia Italy.}}}\ (\bibinfo
  {year} {2010})\ p.\ \bibinfo {pages} {206}\BibitemShut {NoStop}%
\bibitem [{\citenamefont {Carlson}\ and\ \citenamefont
  {Gandolfi}(2014)}]{Carlson2014}%
  \BibitemOpen
  \bibfield  {author} {\bibinfo {author} {\bibfnamefont {J.}~\bibnamefont
  {Carlson}}\ and\ \bibinfo {author} {\bibfnamefont {S.}~\bibnamefont
  {Gandolfi}},\ }\bibfield  {title} {\bibinfo {title} {\emph {Predicting
  energies of small clusters from the inhomogeneous unitary Fermi gas}},\
  }\href {\doibase 10.1103/PhysRevA.90.011601} {\bibfield  {journal} {\bibinfo
  {journal} {Phys. Rev. A}\ }\textbf {\bibinfo {volume} {90}},\ \bibinfo
  {pages} {011601} (\bibinfo {year} {2014})}\BibitemShut {NoStop}%
\bibitem [{\citenamefont {Blume}(2008)}]{Blume2008}%
  \BibitemOpen
  \bibfield  {author} {\bibinfo {author} {\bibfnamefont {D.}~\bibnamefont
  {Blume}},\ }\bibfield  {title} {\bibinfo {title} {\emph {Trapped polarized
  Fermi gas at unitarity}},\ }\href {\doibase 10.1103/PhysRevA.78.013635}
  {\bibfield  {journal} {\bibinfo  {journal} {Phys. Rev. A}\ }\textbf {\bibinfo
  {volume} {78}},\ \bibinfo {pages} {013635} (\bibinfo {year}
  {2008})}\BibitemShut {NoStop}%
\bibitem [{\citenamefont {Carlson}\ \emph {et~al.}(2011)\citenamefont
  {Carlson}, \citenamefont {Gandolfi}, \citenamefont {Schmidt},\ and\
  \citenamefont {Zhang}}]{Carlson2011}%
  \BibitemOpen
  \bibfield  {author} {\bibinfo {author} {\bibfnamefont {J.}~\bibnamefont
  {Carlson}}, \bibinfo {author} {\bibfnamefont {S.}~\bibnamefont {Gandolfi}},
  \bibinfo {author} {\bibfnamefont {K.~E.}\ \bibnamefont {Schmidt}}, \ and\
  \bibinfo {author} {\bibfnamefont {S.}~\bibnamefont {Zhang}},\ }\bibfield
  {title} {\bibinfo {title} {\emph {Auxiliary-field quantum Monte Carlo method
  for strongly paired fermions}},\ }\href {\doibase 10.1103/PhysRevA.84.061602}
  {\bibfield  {journal} {\bibinfo  {journal} {Phys. Rev. A}\ }\textbf {\bibinfo
  {volume} {84}},\ \bibinfo {pages} {061602} (\bibinfo {year}
  {2011})}\BibitemShut {NoStop}%
\bibitem [{\citenamefont {{Levinsen}}\ \emph {et~al.}(2015)\citenamefont
  {{Levinsen}}, \citenamefont {{Massignan}}, \citenamefont {{Bruun}},\ and\
  \citenamefont {{Parish}}}]{Levinsen2015}%
  \BibitemOpen
  \bibfield  {author} {\bibinfo {author} {\bibfnamefont {J.}~\bibnamefont
  {{Levinsen}}}, \bibinfo {author} {\bibfnamefont {P.}~\bibnamefont
  {{Massignan}}}, \bibinfo {author} {\bibfnamefont {G.~M.}\ \bibnamefont
  {{Bruun}}}, \ and\ \bibinfo {author} {\bibfnamefont {M.~M.}\ \bibnamefont
  {{Parish}}},\ }\bibfield  {title} {\bibinfo {title} {\emph {Strong-coupling
  ansatz for the one-dimensional Fermi gas in a harmonic potential}},\ }\href
  {\doibase 10.1126/sciadv.1500197} {\bibfield  {journal} {\bibinfo  {journal}
  {Science Advances}\ }\textbf {\bibinfo {volume} {1}},\ \bibinfo {pages}
  {e1500197} (\bibinfo {year} {2015})}\BibitemShut {NoStop}%
\bibitem [{\citenamefont {Recati}\ \emph {et~al.}(2008)\citenamefont {Recati},
  \citenamefont {Lobo},\ and\ \citenamefont {Stringari}}]{Recati2008}%
  \BibitemOpen
  \bibfield  {author} {\bibinfo {author} {\bibfnamefont {A.}~\bibnamefont
  {Recati}}, \bibinfo {author} {\bibfnamefont {C.}~\bibnamefont {Lobo}}, \ and\
  \bibinfo {author} {\bibfnamefont {S.}~\bibnamefont {Stringari}},\ }\bibfield
  {title} {\bibinfo {title} {\emph {Role of interactions in spin-polarized
  atomic Fermi gases at unitarity}},\ }\href {\doibase
  10.1103/PhysRevA.78.023633} {\bibfield  {journal} {\bibinfo  {journal} {Phys.
  Rev. A}\ }\textbf {\bibinfo {volume} {78}},\ \bibinfo {pages} {023633}
  (\bibinfo {year} {2008})}\BibitemShut {NoStop}%
\bibitem [{\citenamefont {Tan}(2008{\natexlab{a}})}]{Tan2008}%
  \BibitemOpen
  \bibfield  {author} {\bibinfo {author} {\bibfnamefont {S.}~\bibnamefont
  {Tan}},\ }\bibfield  {title} {\bibinfo {title} {\emph {Energetics of a
  strongly correlated Fermi gas}},\ }\href {\doibase
  http://dx.doi.org/10.1016/j.aop.2008.03.004} {\bibfield  {journal} {\bibinfo
  {journal} {Annals of Physics}\ }\textbf {\bibinfo {volume} {323}},\ \bibinfo
  {pages} {2952 } (\bibinfo {year} {2008}{\natexlab{a}})}\BibitemShut {NoStop}%
\bibitem [{\citenamefont {Tan}(2008{\natexlab{b}})}]{Tan2008b}%
  \BibitemOpen
  \bibfield  {author} {\bibinfo {author} {\bibfnamefont {S.}~\bibnamefont
  {Tan}},\ }\bibfield  {title} {\bibinfo {title} {\emph {Large momentum part of
  a strongly correlated Fermi gas}},\ }\href {\doibase
  http://dx.doi.org/10.1016/j.aop.2008.03.005} {\bibfield  {journal} {\bibinfo
  {journal} {Annals of Physics}\ }\textbf {\bibinfo {volume} {323}},\ \bibinfo
  {pages} {2971 } (\bibinfo {year} {2008}{\natexlab{b}})}\BibitemShut {NoStop}%
\bibitem [{\citenamefont {Braaten}(2012)}]{Braaten2012}%
  \BibitemOpen
  \bibfield  {author} {\bibinfo {author} {\bibfnamefont {E.}~\bibnamefont
  {Braaten}},\ }\bibinfo {title} {\emph {Universal Relations for Fermions with
  Large Scattering Length}},\ in\ \href {\doibase 10.1007/978-3-642-21978-8_6}
  {\emph {\bibinfo {booktitle} {The BCS-BEC Crossover and the Unitary Fermi
  Gas}}},\ \bibinfo {editor} {edited by\ \bibinfo {editor} {\bibfnamefont
  {W.}~\bibnamefont {Zwerger}}}\ (\bibinfo  {publisher} {Springer Berlin
  Heidelberg},\ \bibinfo {address} {Berlin, Heidelberg},\ \bibinfo {year}
  {2012})\ pp.\ \bibinfo {pages} {193--231}\BibitemShut {NoStop}%
\bibitem [{\citenamefont {Sagi}\ \emph {et~al.}(2012)\citenamefont {Sagi},
  \citenamefont {Drake}, \citenamefont {Paudel},\ and\ \citenamefont
  {Jin}}]{Sagi2012}%
  \BibitemOpen
  \bibfield  {author} {\bibinfo {author} {\bibfnamefont {Y.}~\bibnamefont
  {Sagi}}, \bibinfo {author} {\bibfnamefont {T.~E.}\ \bibnamefont {Drake}},
  \bibinfo {author} {\bibfnamefont {R.}~\bibnamefont {Paudel}}, \ and\ \bibinfo
  {author} {\bibfnamefont {D.~S.}\ \bibnamefont {Jin}},\ }\bibfield  {title}
  {\bibinfo {title} {\emph {Measurement of the Homogeneous Contact of a Unitary
  Fermi Gas}},\ }\href {\doibase 10.1103/PhysRevLett.109.220402} {\bibfield
  {journal} {\bibinfo  {journal} {Phys. Rev. Lett.}\ }\textbf {\bibinfo
  {volume} {109}},\ \bibinfo {pages} {220402} (\bibinfo {year}
  {2012})}\BibitemShut {NoStop}%
\bibitem [{\citenamefont {Combescot}\ \emph {et~al.}(2006)\citenamefont
  {Combescot}, \citenamefont {Giorgini},\ and\ \citenamefont
  {Stringari}}]{Combescot2006}%
  \BibitemOpen
  \bibfield  {author} {\bibinfo {author} {\bibfnamefont {R.}~\bibnamefont
  {Combescot}}, \bibinfo {author} {\bibfnamefont {S.}~\bibnamefont {Giorgini}},
  \ and\ \bibinfo {author} {\bibfnamefont {S.}~\bibnamefont {Stringari}},\
  }\bibfield  {title} {\bibinfo {title} {\emph {Molecular signatures in the
  structure factor of an interacting Fermi gas}},\ }\href
  {http://stacks.iop.org/0295-5075/75/i=5/a=695} {\bibfield  {journal}
  {\bibinfo  {journal} {EPL (Europhysics Letters)}\ }\textbf {\bibinfo {volume}
  {75}},\ \bibinfo {pages} {695} (\bibinfo {year} {2006})}\BibitemShut
  {NoStop}%
\bibitem [{\citenamefont {Haussmann}\ \emph {et~al.}(2009)\citenamefont
  {Haussmann}, \citenamefont {Punk},\ and\ \citenamefont
  {Zwerger}}]{Haussmann2009}%
  \BibitemOpen
  \bibfield  {author} {\bibinfo {author} {\bibfnamefont {R.}~\bibnamefont
  {Haussmann}}, \bibinfo {author} {\bibfnamefont {M.}~\bibnamefont {Punk}}, \
  and\ \bibinfo {author} {\bibfnamefont {W.}~\bibnamefont {Zwerger}},\
  }\bibfield  {title} {\bibinfo {title} {\emph {Spectral functions and rf
  response of ultracold fermionic atoms}},\ }\href {\doibase
  10.1103/PhysRevA.80.063612} {\bibfield  {journal} {\bibinfo  {journal} {Phys.
  Rev. A}\ }\textbf {\bibinfo {volume} {80}},\ \bibinfo {pages} {063612}
  (\bibinfo {year} {2009})}\BibitemShut {NoStop}%
\bibitem [{\citenamefont {Gandolfi}\ \emph {et~al.}(2011)\citenamefont
  {Gandolfi}, \citenamefont {Schmidt},\ and\ \citenamefont
  {Carlson}}]{Gandolfi2011}%
  \BibitemOpen
  \bibfield  {author} {\bibinfo {author} {\bibfnamefont {S.}~\bibnamefont
  {Gandolfi}}, \bibinfo {author} {\bibfnamefont {K.~E.}\ \bibnamefont
  {Schmidt}}, \ and\ \bibinfo {author} {\bibfnamefont {J.}~\bibnamefont
  {Carlson}},\ }\bibfield  {title} {\bibinfo {title} {\emph {BEC-BCS crossover
  and universal relations in unitary Fermi gases}},\ }\href {\doibase
  10.1103/PhysRevA.83.041601} {\bibfield  {journal} {\bibinfo  {journal} {Phys.
  Rev. A}\ }\textbf {\bibinfo {volume} {83}},\ \bibinfo {pages} {041601}
  (\bibinfo {year} {2011})}\BibitemShut {NoStop}%
\bibitem [{\citenamefont {Palestini}\ \emph {et~al.}(2010)\citenamefont
  {Palestini}, \citenamefont {Perali}, \citenamefont {Pieri},\ and\
  \citenamefont {Strinati}}]{Palestini2010}%
  \BibitemOpen
  \bibfield  {author} {\bibinfo {author} {\bibfnamefont {F.}~\bibnamefont
  {Palestini}}, \bibinfo {author} {\bibfnamefont {A.}~\bibnamefont {Perali}},
  \bibinfo {author} {\bibfnamefont {P.}~\bibnamefont {Pieri}}, \ and\ \bibinfo
  {author} {\bibfnamefont {G.~C.}\ \bibnamefont {Strinati}},\ }\bibfield
  {title} {\bibinfo {title} {\emph {Temperature and coupling dependence of the
  universal contact intensity for an ultracold Fermi gas}},\ }\href {\doibase
  10.1103/PhysRevA.82.021605} {\bibfield  {journal} {\bibinfo  {journal} {Phys.
  Rev. A}\ }\textbf {\bibinfo {volume} {82}},\ \bibinfo {pages} {021605}
  (\bibinfo {year} {2010})}\BibitemShut {NoStop}%
\bibitem [{\citenamefont {Hu}\ \emph {et~al.}(2011)\citenamefont {Hu},
  \citenamefont {Liu},\ and\ \citenamefont {Drummond}}]{Hu2011}%
  \BibitemOpen
  \bibfield  {author} {\bibinfo {author} {\bibfnamefont {H.}~\bibnamefont
  {Hu}}, \bibinfo {author} {\bibfnamefont {X.-J.}\ \bibnamefont {Liu}}, \ and\
  \bibinfo {author} {\bibfnamefont {P.~D.}\ \bibnamefont {Drummond}},\
  }\bibfield  {title} {\bibinfo {title} {\emph {Universal contact of strongly
  interacting fermions at finite temperatures}},\ }\href
  {http://stacks.iop.org/1367-2630/13/i=3/a=035007} {\bibfield  {journal}
  {\bibinfo  {journal} {New Journal of Physics}\ }\textbf {\bibinfo {volume}
  {13}},\ \bibinfo {pages} {035007} (\bibinfo {year} {2011})}\BibitemShut
  {NoStop}%
\bibitem [{\citenamefont {Yan}\ and\ \citenamefont {Blume}(2013)}]{Yan2013}%
  \BibitemOpen
  \bibfield  {author} {\bibinfo {author} {\bibfnamefont {Y.}~\bibnamefont
  {Yan}}\ and\ \bibinfo {author} {\bibfnamefont {D.}~\bibnamefont {Blume}},\
  }\bibfield  {title} {\bibinfo {title} {\emph {Harmonically trapped Fermi gas:
  Temperature dependence of the Tan contact}},\ }\href {\doibase
  10.1103/PhysRevA.88.023616} {\bibfield  {journal} {\bibinfo  {journal} {Phys.
  Rev. A}\ }\textbf {\bibinfo {volume} {88}},\ \bibinfo {pages} {023616}
  (\bibinfo {year} {2013})}\BibitemShut {NoStop}%
\bibitem [{Blu()}]{BlumePC}%
  \BibitemOpen
  \href@noop {} {}\bibinfo {note} {D.~Blume, private communication}\BibitemShut
  {NoStop}%
\bibitem [{\citenamefont {Punk}\ \emph {et~al.}(2009)\citenamefont {Punk},
  \citenamefont {Dumitrescu},\ and\ \citenamefont {Zwerger}}]{Punk2009}%
  \BibitemOpen
  \bibfield  {author} {\bibinfo {author} {\bibfnamefont {M.}~\bibnamefont
  {Punk}}, \bibinfo {author} {\bibfnamefont {P.~T.}\ \bibnamefont
  {Dumitrescu}}, \ and\ \bibinfo {author} {\bibfnamefont {W.}~\bibnamefont
  {Zwerger}},\ }\bibfield  {title} {\bibinfo {title} {\emph
  {Polaron-to-molecule transition in a strongly imbalanced Fermi gas}},\ }\href
  {\doibase 10.1103/PhysRevA.80.053605} {\bibfield  {journal} {\bibinfo
  {journal} {Phys. Rev. A}\ }\textbf {\bibinfo {volume} {80}},\ \bibinfo
  {pages} {053605} (\bibinfo {year} {2009})}\BibitemShut {NoStop}%
\bibitem [{\citenamefont {Kohstall}\ \emph {et~al.}(2012)\citenamefont
  {Kohstall}, \citenamefont {Zaccanti}, \citenamefont {Jag}, \citenamefont
  {Trenkwalder}, \citenamefont {Massignan}, \citenamefont {Bruun},
  \citenamefont {Schreck},\ and\ \citenamefont {Grimm}}]{Kohstall2012}%
  \BibitemOpen
  \bibfield  {author} {\bibinfo {author} {\bibfnamefont {C.}~\bibnamefont
  {Kohstall}}, \bibinfo {author} {\bibfnamefont {M.}~\bibnamefont {Zaccanti}},
  \bibinfo {author} {\bibfnamefont {M.}~\bibnamefont {Jag}}, \bibinfo {author}
  {\bibfnamefont {A.}~\bibnamefont {Trenkwalder}}, \bibinfo {author}
  {\bibfnamefont {P.}~\bibnamefont {Massignan}}, \bibinfo {author}
  {\bibfnamefont {G.~M.}\ \bibnamefont {Bruun}}, \bibinfo {author}
  {\bibfnamefont {F.}~\bibnamefont {Schreck}}, \ and\ \bibinfo {author}
  {\bibfnamefont {R.}~\bibnamefont {Grimm}},\ }\bibfield  {title} {\bibinfo
  {title} {\emph {Metastability and coherence of repulsive polarons in a
  strongly interacting Fermi mixture}},\ }\href
  {http://dx.doi.org/10.1038/nature11065} {\bibfield  {journal} {\bibinfo
  {journal} {Nature}\ }\textbf {\bibinfo {volume} {485}},\ \bibinfo {pages}
  {615} (\bibinfo {year} {2012})}\BibitemShut {NoStop}%
\bibitem [{\citenamefont {Parish}\ and\ \citenamefont
  {Levinsen}(2016)}]{Parish2016}%
  \BibitemOpen
  \bibfield  {author} {\bibinfo {author} {\bibfnamefont {M.~M.}\ \bibnamefont
  {Parish}}\ and\ \bibinfo {author} {\bibfnamefont {J.}~\bibnamefont
  {Levinsen}},\ }\bibfield  {title} {\bibinfo {title} {\emph {Quantum dynamics
  of impurities coupled to a Fermi sea}},\ }\href {\doibase
  10.1103/PhysRevB.94.184303} {\bibfield  {journal} {\bibinfo  {journal} {Phys.
  Rev. B}\ }\textbf {\bibinfo {volume} {94}},\ \bibinfo {pages} {184303}
  (\bibinfo {year} {2016})}\BibitemShut {NoStop}%
\bibitem [{\citenamefont {{Scazza}}\ \emph {et~al.}(2016)\citenamefont
  {{Scazza}}, \citenamefont {{Valtolina}}, \citenamefont {{Massignan}},
  \citenamefont {{Recati}}, \citenamefont {{Amico}}, \citenamefont
  {{Burchianti}}, \citenamefont {{Fort}}, \citenamefont {{Inguscio}},
  \citenamefont {{Zaccanti}},\ and\ \citenamefont {{Roati}}}]{Scazza2016}%
  \BibitemOpen
  \bibfield  {author} {\bibinfo {author} {\bibfnamefont {F.}~\bibnamefont
  {{Scazza}}}, \bibinfo {author} {\bibfnamefont {G.}~\bibnamefont
  {{Valtolina}}}, \bibinfo {author} {\bibfnamefont {P.}~\bibnamefont
  {{Massignan}}}, \bibinfo {author} {\bibfnamefont {A.}~\bibnamefont
  {{Recati}}}, \bibinfo {author} {\bibfnamefont {A.}~\bibnamefont {{Amico}}},
  \bibinfo {author} {\bibfnamefont {A.}~\bibnamefont {{Burchianti}}}, \bibinfo
  {author} {\bibfnamefont {C.}~\bibnamefont {{Fort}}}, \bibinfo {author}
  {\bibfnamefont {M.}~\bibnamefont {{Inguscio}}}, \bibinfo {author}
  {\bibfnamefont {M.}~\bibnamefont {{Zaccanti}}}, \ and\ \bibinfo {author}
  {\bibfnamefont {G.}~\bibnamefont {{Roati}}},\ }\bibfield  {title} {\bibinfo
  {title} {\emph {{Observation of repulsive Fermi polarons in a resonant
  mixture of ultracold ${}^6$Li atoms}}},\ }\href
  {https://arxiv.org/abs/1609.09817} {\bibfield  {journal} {\bibinfo  {journal}
  {arXiv:1609.09817}\ } (\bibinfo {year} {2016})}\BibitemShut {NoStop}%
\bibitem [{\citenamefont {Schirotzek}\ \emph {et~al.}(2009)\citenamefont
  {Schirotzek}, \citenamefont {Wu}, \citenamefont {Sommer},\ and\ \citenamefont
  {Zwierlein}}]{Schirotzek2009}%
  \BibitemOpen
  \bibfield  {author} {\bibinfo {author} {\bibfnamefont {A.}~\bibnamefont
  {Schirotzek}}, \bibinfo {author} {\bibfnamefont {C.-H.}\ \bibnamefont {Wu}},
  \bibinfo {author} {\bibfnamefont {A.}~\bibnamefont {Sommer}}, \ and\ \bibinfo
  {author} {\bibfnamefont {M.~W.}\ \bibnamefont {Zwierlein}},\ }\bibfield
  {title} {\bibinfo {title} {\emph {Observation of Fermi Polarons in a Tunable
  Fermi Liquid of Ultracold Atoms}},\ }\href {\doibase
  10.1103/PhysRevLett.102.230402} {\bibfield  {journal} {\bibinfo  {journal}
  {Phys. Rev. Lett.}\ }\textbf {\bibinfo {volume} {102}},\ \bibinfo {pages}
  {230402} (\bibinfo {year} {2009})}\BibitemShut {NoStop}%
\bibitem [{\citenamefont {Gaunt}\ \emph {et~al.}(2013)\citenamefont {Gaunt},
  \citenamefont {Schmidutz}, \citenamefont {Gotlibovych}, \citenamefont
  {Smith},\ and\ \citenamefont {Hadzibabic}}]{PhysRevLett.110.200406}%
  \BibitemOpen
  \bibfield  {author} {\bibinfo {author} {\bibfnamefont {A.~L.}\ \bibnamefont
  {Gaunt}}, \bibinfo {author} {\bibfnamefont {T.~F.}\ \bibnamefont
  {Schmidutz}}, \bibinfo {author} {\bibfnamefont {I.}~\bibnamefont
  {Gotlibovych}}, \bibinfo {author} {\bibfnamefont {R.~P.}\ \bibnamefont
  {Smith}}, \ and\ \bibinfo {author} {\bibfnamefont {Z.}~\bibnamefont
  {Hadzibabic}},\ }\bibfield  {title} {\bibinfo {title} {\emph {Bose-Einstein
  Condensation of Atoms in a Uniform Potential}},\ }\href {\doibase
  10.1103/PhysRevLett.110.200406} {\bibfield  {journal} {\bibinfo  {journal}
  {Phys. Rev. Lett.}\ }\textbf {\bibinfo {volume} {110}},\ \bibinfo {pages}
  {200406} (\bibinfo {year} {2013})}\BibitemShut {NoStop}%
\bibitem [{\citenamefont {Efimov}(1970)}]{Efimov1970a}%
  \BibitemOpen
  \bibfield  {author} {\bibinfo {author} {\bibfnamefont {V.}~\bibnamefont
  {Efimov}},\ }\bibfield  {title} {\bibinfo {title} {\emph {Weakly-bound states
  of three resonantly-interacting particles}},\ }\href@noop {} {\bibfield
  {journal} {\bibinfo  {journal} {Yad. Fiz.}\ }\textbf {\bibinfo {volume}
  {12}},\ \bibinfo {pages} {1080} (\bibinfo {year} {1970})},\ \bibinfo {note}
  {[Sov. J. Nucl. Phys. 12, 589-595 (1971)]}\BibitemShut {NoStop}%
\bibitem [{\citenamefont {Braaten}\ and\ \citenamefont
  {Hammer}(2006)}]{braaten2006universality}%
  \BibitemOpen
  \bibfield  {author} {\bibinfo {author} {\bibfnamefont {E.}~\bibnamefont
  {Braaten}}\ and\ \bibinfo {author} {\bibfnamefont {H.-W.}\ \bibnamefont
  {Hammer}},\ }\bibfield  {title} {\bibinfo {title} {\emph {Universality in
  few-body systems with large scattering length}},\ }\href {\doibase
  http://dx.doi.org/10.1016/j.physrep.2006.03.001} {\bibfield  {journal}
  {\bibinfo  {journal} {Physics Reports}\ }\textbf {\bibinfo {volume} {428}},\
  \bibinfo {pages} {259} (\bibinfo {year} {2006})}\BibitemShut {NoStop}%
\bibitem [{\citenamefont {Ferlaino}\ and\ \citenamefont
  {Grimm}(2010)}]{ferlaino2010trend}%
  \BibitemOpen
  \bibfield  {author} {\bibinfo {author} {\bibfnamefont {F.}~\bibnamefont
  {Ferlaino}}\ and\ \bibinfo {author} {\bibfnamefont {R.}~\bibnamefont
  {Grimm}},\ }\bibfield  {title} {\bibinfo {title} {\emph {Trend: Forty years
  of Efimov physics: How a bizarre prediction turned into a hot topic}},\
  }\href {\doibase 10.1103/Physics.3.9} {\bibfield  {journal} {\bibinfo
  {journal} {Physics}\ }\textbf {\bibinfo {volume} {3}},\ \bibinfo {pages} {9}
  (\bibinfo {year} {2010})}\BibitemShut {NoStop}%
\bibitem [{\citenamefont {Naidon}\ and\ \citenamefont
  {Endo}(2016)}]{naidon2016efimov}%
  \BibitemOpen
  \bibfield  {author} {\bibinfo {author} {\bibfnamefont {P.}~\bibnamefont
  {Naidon}}\ and\ \bibinfo {author} {\bibfnamefont {S.}~\bibnamefont {Endo}},\
  }\bibfield  {title} {\bibinfo {title} {\emph {Efimov Physics: a review}},\
  }\href {https://arxiv.org/abs/1610.09805} {\bibfield  {journal} {\bibinfo
  {journal} {arXiv:1610.09805}\ } (\bibinfo {year} {2016})}\BibitemShut
  {NoStop}%
\bibitem [{\citenamefont {Efimov}(1973)}]{Efimov1973}%
  \BibitemOpen
  \bibfield  {author} {\bibinfo {author} {\bibfnamefont {V.}~\bibnamefont
  {Efimov}},\ }\bibfield  {title} {\bibinfo {title} {\emph {Energy levels of
  three resonantly interacting particles}},\ }\href {\doibase
  http://dx.doi.org/10.1016/0375-9474(73)90510-1} {\bibfield  {journal}
  {\bibinfo  {journal} {Nuclear Physics A}\ }\textbf {\bibinfo {volume}
  {210}},\ \bibinfo {pages} {157 } (\bibinfo {year} {1973})}\BibitemShut
  {NoStop}%
\bibitem [{\citenamefont {Castin}\ \emph {et~al.}(2010)\citenamefont {Castin},
  \citenamefont {Mora},\ and\ \citenamefont {Pricoupenko}}]{Castin2010}%
  \BibitemOpen
  \bibfield  {author} {\bibinfo {author} {\bibfnamefont {Y.}~\bibnamefont
  {Castin}}, \bibinfo {author} {\bibfnamefont {C.}~\bibnamefont {Mora}}, \ and\
  \bibinfo {author} {\bibfnamefont {L.}~\bibnamefont {Pricoupenko}},\
  }\bibfield  {title} {\bibinfo {title} {\emph {Four-Body Efimov Effect for
  Three Fermions and a Lighter Particle}},\ }\href {\doibase
  10.1103/PhysRevLett.105.223201} {\bibfield  {journal} {\bibinfo  {journal}
  {Phys. Rev. Lett.}\ }\textbf {\bibinfo {volume} {105}},\ \bibinfo {pages}
  {223201} (\bibinfo {year} {2010})}\BibitemShut {NoStop}%
\bibitem [{\citenamefont {Moser}\ and\ \citenamefont
  {Seiringer}(2016)}]{moser2016stability}%
  \BibitemOpen
  \bibfield  {author} {\bibinfo {author} {\bibfnamefont {T.}~\bibnamefont
  {Moser}}\ and\ \bibinfo {author} {\bibfnamefont {R.}~\bibnamefont
  {Seiringer}},\ }\bibfield  {title} {\bibinfo {title} {\emph {Stability of a
  fermionic $ N+ 1$ particle system with point interactions}},\ }\href
  {https://arxiv.org/abs/1609.08342} {\bibfield  {journal} {\bibinfo  {journal}
  {arXiv:1609.08342}\ } (\bibinfo {year} {2016})}\BibitemShut {NoStop}%
\end{thebibliography}%

\end{document}